\documentclass[a4paper]{article}
\usepackage{moreverb,url}
\usepackage[numbers]{natbib}
\usepackage{amsmath, amssymb}
\usepackage[margin=2.5cm]{geometry}
\usepackage{authblk}
\usepackage{graphicx}
\usepackage{multirow}
\usepackage{multicol}
\usepackage{longtable}
\usepackage{dsfont}
\usepackage{enumitem}
\usepackage{tikz}
\usepackage{microtype}
\usetikzlibrary{positioning, shapes.geometric}
\usepackage{booktabs, tabularx}
\usepackage{lscape}
\usepackage{float}
\usepackage{changepage}
\usepackage[dvipsnames]{xcolor}
\usepackage{listings}
\usepackage{array}
\newcolumntype{C}[1]{>{\centering\arraybackslash}p{#1}}

\usepackage[english]{babel}
\usepackage{csquotes}

\usepackage{nicefrac}
\usepackage[colorlinks,bookmarksopen,bookmarksnumbered,citecolor=blue,urlcolor=blue, linkcolor=blue]{hyperref}
\newcommand{\keywords}[1]{\par\addvspace\baselineskip\noindent\textbf{Keywords:}\enspace #1}

\author[1]{Johannes Vilsmeier}
\author[1]{Fabian Eibensteiner}
\author[1]{Franz König}
\author[2]{Francois Mercier}
\author[1]{Robin Ristl}
\author[3]{Nigel Stallard}
\author[4,5]{Marc Vandemeulebroecke}
\author[6]{Sarah Zohar}
\author[1]{Martin Posch\thanks{Corresponding author: \url{martin.posch@meduniwien.ac.at}}}

\affil[1]{Institute of Medical Statistics, Center for Medical Data Science, Medical University of Vienna}
\affil[2]{Hoffmann-La Roche AG, gRED Modeling \& Simulation}
\affil[3]{Warwick Clinical Trials Unit, University of Warwick}
\affil[4]{Radboud University Medical Center}
\affil[5]{Bayer BCC, Basel, Switzerland}
\affil[6]{Inserm, Université Paris Cité, Inria, HeKA}

\date{}

\title{Clustering-Based Outcome Models for Clinical Studies: A Scoping Review}

\begin{document}

\maketitle

\begin{abstract}
This review provides a systematic overview of methods that combine 
covariate-based clustering of observational units (patients) with outcome models for 
clinical studies. We distinguish between informed-cluster models, where the outcome contributes to cluster 
formation, and agnostic-cluster models, where clustering is performed 
solely on covariates in a separate first step. Informed-cluster models
include product partition models with covariates (PPMx), finite mixtures 
of regression models (FMR), and cluster-aware supervised learning (CluSL). 
Agnostic-cluster models encompass two-step procedures using either 
model-based or algorithmic clustering followed by cluster-specific 
regression models. Following a systematic search of Web of Science and 
PubMed, 55 records were identified that propose or evaluate such 
models. We describe the key models, summarise study 
characteristics, and present applications from biomedical and public 
health research. Clustering-based outcome models are particularly 
relevant for settings with high-dimensional covariates (e.g., biomarker panels and "omics") and heterogeneous 
patient populations. These models can support risk stratification and we discuss extensions to estimate subgroup-specific treatment effects. They are most valuable when the population is clustered in distinct regions of the covariate space that correspond to different outcome distributions. We discuss applications to rare disease research, covariate adjustment and borrowing from historical data, and subgroup-specific treatment effect estimation in clinical trials.
\end{abstract}

\keywords{Clustering, finite mixture regression, subgroup identification, product partition models, prognostic models, precision medicine, covariate-dependent clustering}

\section{Introduction}

Individuals with the same underlying disease or health condition may differ 
substantially in how they manifest symptoms and how their disease progresses. Some of the resulting  variability in clinical outcomes may be systematic and determined by baseline characteristics of patients.
The systematic part of the variation, termed \emph{prognostic heterogeneity} can be described by models based on measured \emph{prognostic covariates}, i.e., covariates associated with the outcome of interest regardless of which treatment
a patient received  \cite{ballmanBiomarkerPredictivePrognostic2015, sechidisDistinguishingPrognosticPredictive2018}. 
A related concept, \emph{predictive heterogeneity} (or heterogeneity of treatment effects, HTE; see supplementary materials S1 for a list of abbreviations), describes 
inter-individual variability in the effect of a treatment depending on covariate status. Covariates that modify 
treatment effects are termed \emph{predictive covariates} \cite{kentPersonalizedEvidenceBased2018}.

Outcome models incorporating prognostic or predictive covariates  are widely used \cite{arshi2025, rileyPrognosisResearchStrategy2013}
and there is a broad range of methodological research on approaches that account for prognostic or predictive heterogeneity through covariates
\cite{lipkovichModernApproachesEvaluating2024, 
ondraMethodsIdentificationConfirmation2016, 
zhuIdentificationPrognosticPredictive2023, frommletSelectingPredictiveBiomarkers2022, 
lipkovichTutorialBiostatisticsDatadriven2017}. 
These methods cover settings where a small set of candidate covariates is known a priori as well as settings where such covariates are selected in a data-driven manner from a high-dimensional covariate vector. Outcome models include prognostic models as well as models with treatment–covariate interactions to assess treatment effect heterogeneity.

Objectives of outcome modelling are \emph{explanation} or \emph{prediction}  \citep{efronPredictionEstimationAttribution2020}. 
For explanation, the goal is to quantify the relative contribution of each covariate, a task that becomes increasingly 
difficult as the number of covariates and their interactions grows. 
When the goal is prediction, the model fit and generalisability to new observations are the main concern 
rather than to isolate individual effects. In both settings, variable selection approaches are used to obtain 
parsimonious prognostic models and aim to reduce overfitting resulting in poor out-of-sample performance. The selection 
of variables is ideally guided by subject-matter knowledge but can also be data-driven. A wide range of selection 
approaches have been proposed, such as penalised regression methods including the \emph{least absolute shrinkage and selection operator} 
(LASSO) or elastic net approaches (\citep[see e.g., Heinze et al.][for a review and recommendations for best practices]{heinzeVariableSelectionReview2018}).

However, in particular  when the number of candidate covariates is large relative to the sample size, overfitting can undermine 
both objectives and may lead to unstable effect estimates and low predictive accuracy in new observations. This is especially  relevant in rare disease research where cohorts are often small while for each patient a rich set of information such as large biomarker panels may be available.
In such settings, clustering patients can provide a low-dimensional representation of baseline covariate profiles that can capture more complex patterns without the need to  explicitly specify high-order interactions in the outcome model.

In this review we give an overview of published methods that use clustering of observational units (patients) 
based on covariates in the process of building outcome models, for example, regression models.  

Clustering algorithms partition objects into groups based on their characteristics. 
The goal is to form groups such that objects within a group are more similar to each other than to those 
in other groups \citep[see e.g., Izenman][]{izenman2008modern} (Chapter 12). One can cluster the covariates or the observational units to
obtain a lower-dimensional representation of the data. A typical approach to 
dimensionality reduction through clustering of covariates would be to first 
group covariates based on their similarity and then select one covariate as 
a representative for use in the statistical model for the clinical outcome. Such 
approaches have been recently reviewed by Kuzudisli et al. \cite{kuzudisliReviewFeatureSelection2023}. 

This review focuses on  methods 
that cluster observational units. Rather than relating each covariate 
to the clinical outcome, these methods summarise covariate information into cluster membership indicators, dissimilarities to cluster centroids, or other variables that capture similarity among covariate profiles. The proposed 
approaches range from simple two-step procedures where observational units are first 
clustered based on covariate similarity and one outcome model is then fitted within each cluster, to approaches that perform clustering and outcome modelling jointly. 
We refer to these methods as \emph{clustering-based outcome models}.
Depending on the objective, the resulting clusters or subgroups can be either viewed 
as tools in an outcome model to explain heterogeneity or can be interpreted as estimates of biologically distinct
subpopulations within the population from which the sample was drawn.

Prognostic models based on clustering are related to traditional subgroup analysis, where the covariate space is also 
partitioned to obtain more homogeneous within-group estimates of outcomes (or, in interventional settings, treatment effects).
In subgroup analysis, groups are often defined by discretising one or a small number of covariates. Clustering 
follows a similar logic but can form groups from larger covariate sets in a data-driven way. However, if the outcome 
is not used in the estimation of clusters, improved within-group outcome homogeneity can only be expected if the
true underlying subgroup structure in covariate space
corresponds to clusters in the outcome space. 
These approaches are expected to be most efficient when the population consists of distinct subpopulations in the covariate space matching different outcome distributions or treatment effects. 

Clustering-based outcome models can be classified into outcome models with \emph{outcome-informed} clustering and models with \emph{outcome-agnostic} clustering. We refer to these as \emph{informed-cluster models} and \emph{agnostic-cluster models}, respectively. In informed-cluster models, the outcome variable is used for clustering while in agnostic-cluster models, clustering is performed solely based on covariate information and the outcome variable is used only in a subsequent outcome-modelling step. Cluster-derived variables such as cluster membership are used as covariates in a model for the outcome. 

Note that the terms \emph{informed} and \emph{agnostic} only refer to the estimation of clusters. The subsequent prediction models are always supervised and learn mappings from the covariates to the outcome using training data that contain outcome values. The clustering step  itself, however, is unsupervised in both cases because the data sets used to estimate the clusters contain no cluster membership labels.

Including clustering-derived variables in models for clinical outcomes may serve several purposes beyond dimensionality reduction. For instance, augmenting the covariate space with cluster membership indicators or dissimilarities to cluster centroids can improve predictive performance 
\cite{trivediUtilityClusteringPrediction, piernikStudyUsingData2021} by enabling the model to capture  complex covariate-outcome relationships in settings where the outcome differs across well defined discrete subgroups defined by multiple covariates.
In randomized trials, this approach can be used for covariate adjustment and to assess subgroup-specific treatment effects.
Clustering is also an approach to incorporate functional covariates into outcome models, such as irregularly 
spaced measurements over time \cite{luBayesianConsensusClustering2022, garciaUnsupervisedBayesianClassification2024}, 
or variable-length covariates that arise in longitudinal studies where the number of visits differs across patients 
\cite{pageClusteringPredictionVariable2022}.

The performance of the methods depends on the available sample size. Therefore, we report sample sizes and the number of covariates used in the records  to help identify clustering-based prediction approaches that can be applied in smaller data sets, such as in rare disease research where sample sizes are inherently 
small, but heterogeneity in symptom manifestation and treatment response is a central characteristic \cite{morelMeasuringWhatMatters2017, murrayApproachesAssessmentClinical2023}.

In this review we focus on clustering-based outcome models that have been applied to clinical studies but we also 
include methods proposed in engineering and computer science. Records where a method is only applied 
but not described in detail and methods developed  exclusively for imaging data are not covered. We illustrate 
selected methods published in biomedical and public health journals using applications taken from the source publications. 
Finally, we discuss potential fields of application for clustering-based outcome models with a particular 
emphasis on research in rare-diseases and clinical trial designs incorporating historical data.

\section{Methods}
\label{sec:methods}

The complete study protocol detailing the screening and data extraction process was registered with the Open Science Framework and is available at 
\url{https://osf.io/e6k8f}.

\subsection{Search strategy}
\label{sec:search}

Relevant literature was identified by searching Web of Science (WoS) and PubMed using the 
search strings listed in Table~\ref{tbl:search}. In addition, five records 
\citep{leischFlexMixGeneralFramework2004, grunFlexMixVersion22008, proust-limaDevelopmentValidationDynamic2009, trivediUtilityClusteringPrediction, daytonConcomitantVariableLatentClassModels} 
that were not contained in either the PubMed or the WoS search results were added manually.
Records were included if they met two criteria: (1) the record proposes or evaluates a method that uses some form of clustering of observational units based on covariates and 
uses this clustering to model an outcome, and (2) the record provides sufficient details on the method(s), such that
it could be independently implemented based on the provided description.
Records were excluded if they focused solely on clustering methods without outcome modelling, clustered
covariates rather than observational units, presented only applications without methodological contributions, 
analysed imaging data exclusively, introduced software only or were reviews of existing methods.

\subsection{Screening process}

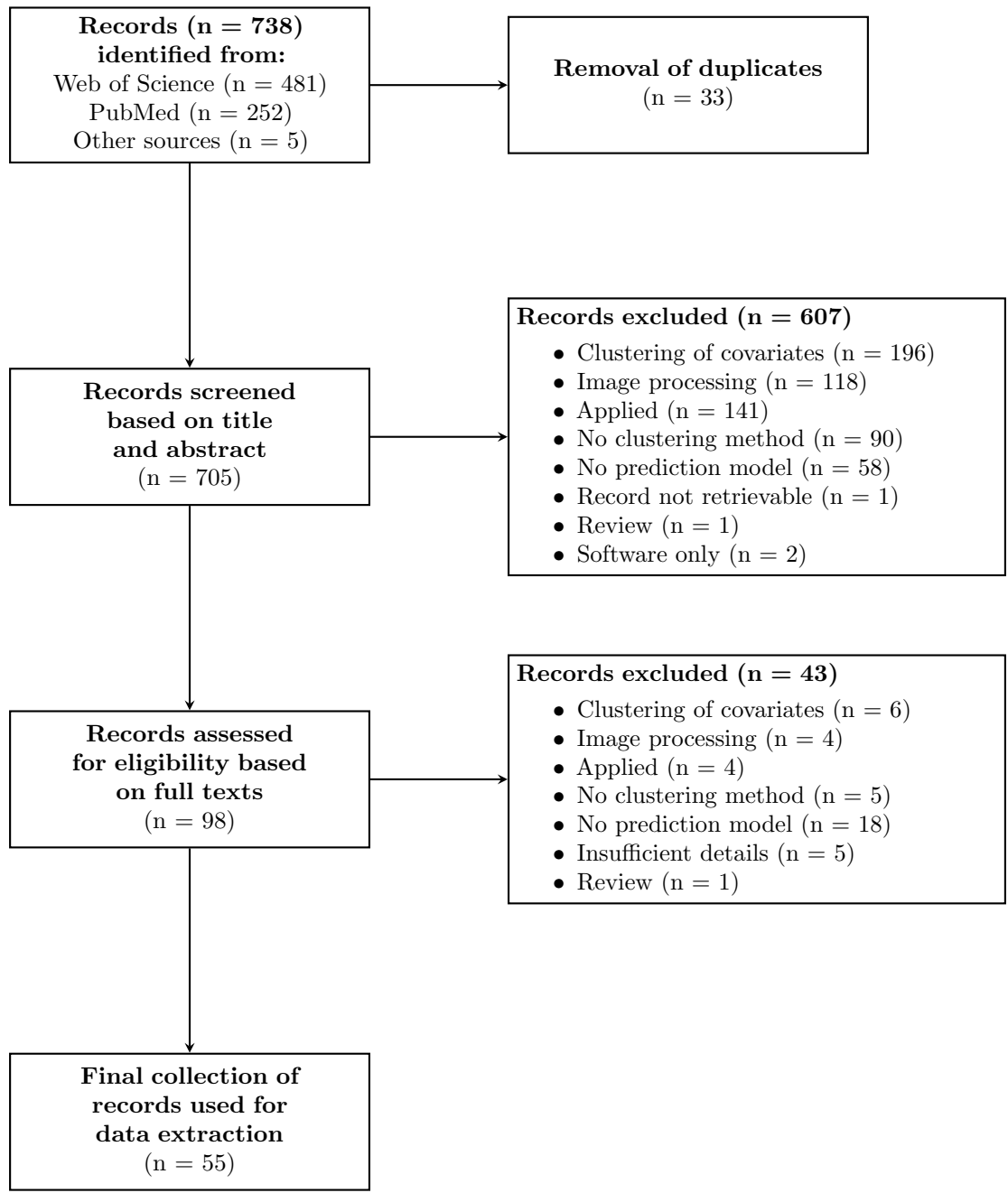
\begin{figure}
  \centering
\tikzset{
    box/.style={
        rectangle,
        draw,
        thick,
        text width=5cm,
        minimum height=2cm,
        align=center,
        font=\normalsize
    },
    excludebox/.style={
        rectangle,
        draw,
        thick,
        text width=7cm,
        minimum height=3.5cm,
        align=left,
        font=\normalsize
    },
    arrow/.style={
        ->,
        thick,
        >=stealth
    }
}
\newcommand{\compactitem}{\setlength{\itemsep}{0pt}\setlength{\parskip}{0pt}}

\begin{tikzpicture}[node distance=3cm]
    \node[box] (identification) {
        \textbf{Records (n = 738) identified from:}\\
        Web of Science (n = 481)\\
        PubMed (n = 252)\\
        Other sources (n = 5)
    };
    
    \node[box, right=2cm of identification] (duplicates) {
        \textbf{Removal of duplicates}\\
        (n = 33)
    };
    
    \node[box, below=of identification] (screening) {
        \textbf{Records screened based on title\\and abstract}\\
        (n = 705)
    };
    
    \node[excludebox, right=2cm of screening] (excluded1) {
        \textbf{Records excluded (n = 607)}
        \vspace{-0.5em}
        \begin{itemize}\compactitem
            \item Clustering of covariates (n = 196)
            \item Image processing (n = 118)
            \item Applied (n = 141)
            \item No clustering method (n = 90)
            \item No prediction model (n = 58)
            \item Record not retrievable (n = 1)
            \item Review (n = 1)
            \item Software only (n = 2)
        \end{itemize}
    };
    
    \node[box, below=of screening] (eligibility) {
        \textbf{Records assessed for eligibility based\\on full texts}\\
        (n = 98)
    };
    
    \node[excludebox, right=2cm of eligibility] (excluded2) {
        \textbf{Records excluded (n = 43)}
        \vspace{-0.5em}
        \begin{itemize}\compactitem
            \item Clustering of covariates (n = 6)
            \item Image processing (n = 4)
            \item Applied (n = 4)
            \item No clustering method (n = 5)
            \item No prediction model (n = 18)
            \item Insufficient details (n = 5)
            \item Review (n = 1)
        \end{itemize}
    };
    
    \node[box, below=of eligibility] (final) {
        \textbf{Final collection of records used for\\data extraction}\\
        (n = 55)
    };
    
    \draw[arrow] (identification) -- (duplicates);
    \draw[arrow] (identification) -- (screening);
    \draw[arrow] (screening) -- (excluded1);
    \draw[arrow] (screening) -- (eligibility);
    \draw[arrow] (eligibility) -- (excluded2);
    \draw[arrow] (eligibility) -- (final);
\end{tikzpicture}

\caption{
PRISMA flow chart showing the literature search and screening process. 
The five records identified through other sources were either cited in included studies or known to the reviewers from prior work.
} 
\label{fig:prisma_flow} 
\end{figure}

The screening process was divided into two stages. First, after removing duplicated records across databases, titles and 
abstracts were screened for inclusion by one researcher (JV). Second, the resulting selection of records was distributed 
equally among eight reviewers, who made final inclusion decisions and extracted information using an extraction form 
(see Section \ref{sec:extraction}) based on full-text reviews. 
To supplement this process, full texts were also uploaded to Elicit \url{https://elicit.com}, a large language model-based 
research assistant designed for systematic literature reviews. Elicit 
was instructed to apply the same inclusion/exclusion criteria and extraction form as the reviewers 
(see supplementary materials S3 for a list of the prompts used). The complete screening process, including the number of 
identified records at each stage, is summarised in Figure~\ref{fig:prisma_flow}. If discrepancies 
between reviewer-based and Elicit-based decisions occurred, selected reviewers discussed these 
discrepancies and reached consensus.

\subsection{Extracted information}\label{sec:extraction}
Reviewers were instructed to extract the following data from each record: \textit{Study characteristics:} Bibliographic information, sample sizes, number of covariates, and domains of the real data used.
\textit{Methodological details:} Type of clustering methods employed, whether the clustering methods are algorithmic or model-based, 
how the number of clusters was determined and software availability.
\textit{Study design and objectives:} Objectives pursued by introducing clustering, types of outcomes the method was developed for, and whether the method is an informed-cluster or an 
agnostic-cluster model.
\textit{Performance evaluation:} Performance measures used in simulation studies and case studies using real-world data. The 
extraction form can be found in Table~S2 in supplement S2 as well as the accompanying online PDF (\url{https://osf.io/e6k8f/files/rsfh4}) and in the supplementary materials S2.

\subsection{Synthesis}

\begin{figure}[!h]
  \centering
\tikzset{
    topbox/.style={
        rectangle,
        draw,
        thick,
        text width=4cm,
        minimum height=1.5cm,
        align=center,
        font=\normalsize
    },
    mainbox/.style={
        rectangle,
        draw,
        thick,
        text width=4.3cm,
        minimum height=1.5cm,
        align=center,
        font=\normalsize
    },
    detailbox/.style={
        rectangle,
        draw,
        thick,
        minimum width=3cm,
        minimum height=1cm,
        align=left,
        font=\normalsize
    },
    arrow/.style={
        ->,
        thick,
        >=stealth
    }
}

\begin{tikzpicture}[node distance=1cm]
    \node[topbox] (top) {
        Records containing\\
        clustering-based outcome models\\
        (n = 55)
    };
    
    \node[mainbox, below left=0.5cm and -1cm of top] (informed) {
        Informed-cluster models\\
        (n = 28)
    };
    
    \node[mainbox, below right=0.5cm and -1cm of top] (agnostic) {
        Agnostic-cluster models\\
        (n = 32)
    };
    
    \node[detailbox, below=of informed] (informeddetail) {
        PPM/PPMx (n = 19)\\
        FMR (n = 8)\\
        CluSL (n = 1)
    };
    
    \node[detailbox, below left=1cm and -2.5cm of agnostic] (algorithmic) {
        \textbf{Algorithmic}\\
        $k$-means clustering (n = 14)\\
        Hierarchical clustering (n = 5)\\
        $k$-medoids clustering (n = 2)\\
        Spectral clustering (n = 2)\\
        Other (n = 5)
    };
    
    \node[detailbox, below right=1cm and -2cm of agnostic] (modelbased) {
        \textbf{Model-based}\\
        Finite mixture model\\
        clustering (n = 4)
    };
    
    \draw[arrow] (top) -| (informed);
    \draw[arrow] (top) -| (agnostic);
    \draw[arrow] (informed) -- (informeddetail);
    \draw[arrow] (agnostic.210) -- (algorithmic.62);
    \draw[arrow] (agnostic.331) -- (modelbased.125);
\end{tikzpicture}
  \caption{
  Categorisation of clustering-based outcome models identified in the review of 55 records. Models are 
  first categorised by whether the outcome variable contributes to cluster formation (informed-cluster models) or 
  whether clustering is performed solely on covariates (agnostic-clustering models). Agnostic-clustering models are further subdivided 
  into algorithmic and model-based clustering approaches. Numbers in parentheses indicate counts in each 
  category; totals exceed 55 because some records reported multiple methods.
  PPM/PPMx = Product partition models/product partition models with covariates; FMR = Finite mixtures of regression models; 
  CluSL = Cluster-aware supervised learning. The category \emph{Other} comprises cosine-similarity clustering \cite{gaoFeatureReductionText2012a}, fuzzy $C$-means clustering and 
  subtractive clustering \cite{poojamrHybridDecisionSupport2015}, and a three-step clustering approach for 
  longitudinal data \cite{nguyenMultivariateLongitudinalData2023}. 
}
  \label{fig:classification}
\end{figure}
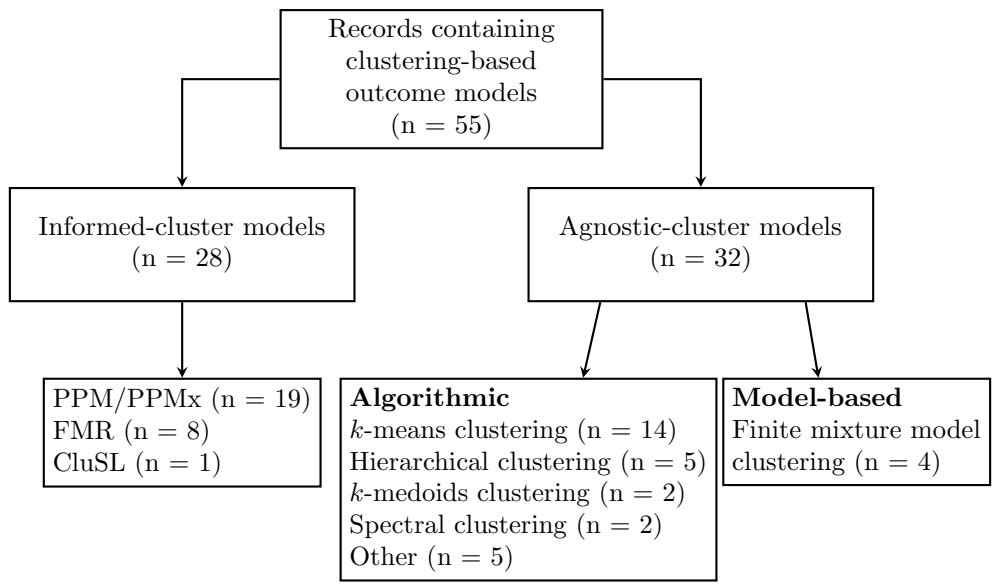

In presenting the identified methods we distinguish primarily between informed-cluster models and agnostic-cluster models (see Table~\ref{tbl:overview_methods} and Figure~\ref{fig:classification}). In informed-cluster 
models, the outcome contributes to the process of finding the clusters. Agnostic-cluster models define clusters solely from covariates 
in a first step, incorporating them into a model for the outcome in a second step. Finite mixtures of regression models \cite{grunFlexMixVersion22008}(FMR) provide typical examples of informed cluster models. Here, a finite mixture distribution is assumed for the outcome 
with the mixing probabilities depending on the covariates (see Section~\ref{sec:FMR}). As an example of an agnostic-cluster model, 
Trivedi et al. \cite{trivediUtilityClusteringPrediction} applied $k$-means clustering of observational units 
based on covariates in a first step, fitted separate cluster-specific linear regression 
models in a second step, repeated this process for several cluster sizes and combined the 
resulting predictions into an ensemble prediction by averaging. Table~\ref{tbl:overview_methods} presents characteristics of informed-cluster models and 
agnostic-cluster models, along with examples and references for further details.

\begin{table}
  \caption{\enspace Overview of informed-cluster and agnostic-cluster models.
    PPM/PPMx = Product partition models / product partition models with covariates, 
    FMR = Finite mixtures of regression,
    CluSL = Cluster-aware supervised learning 
  }
  \label{tbl:overview_methods}
  \centering 
  \begin{tabular}{p{2.3cm}p{11cm}}
    \toprule
    Method & Description \\
    \midrule 
    \textbf{Informed-cluster} &
    Both the outcome and the covariates contribute to the formation of clusters. 
    Cluster-specific outcome models and cluster assignments are estimated jointly. \\ 
    \addlinespace[6pt]
    PPM/PPMx (Section~\ref{sec:PPMx}) & 
    The number of clusters is random, with prior distribution over all possible partitions of the sample depending on 
    a cohesion function and a covariate-similarity function. References: 
    \cite{argientoClusteringBloodDonors2024, barcellaBayesianNonparametricModel2016, barcellaVariableSelectionCovariate2016, jordanStatisticalModellingUsing2007, kyungBayesianAnalysisRandom2017, leiJointAnalysisTwo2025, lijoiPitmanYorMultinomial2020, martinez-vargasPottsCoxSurvivalRegression2023, mullerBayesianInferenceLongitudinal2014, mullerProductPartitionModel2011, muruaSemiparametricBayesianRegression2017, pageCalibratingCovariateInformed2018, pageClusteringPredictionVariable2022, pageDiscoveringInteractionsUsing2021, pagePredictionsBasedClustering2015, quintanaClusterSpecificVariableSelection2015, sunDirichletProcessMixture2017, wadeImprovingPredictionDirichlet, heinerProjectionApproachLocal2025}. \\ 
    \addlinespace[6pt]
    FMR models (Section~\ref{sec:FMR}) & 
    The number of clusters is usually fixed (e.g., determined by some model-selection criterion such as the BIC). Cluster assignment probabilities are modelled explicitly as functions of covariates. References:
    \cite{daytonConcomitantVariableLatentClassModels, garciaUnsupervisedBayesianClassification2024, grunFlexMixVersion22008, leischFlexMixGeneralFramework2004, proust-limaDevelopmentValidationDynamic2009, proust-limaJointLatentClass2014, wangMixturesT2024, zhangJointLatentClass2022}. \\ 
    \addlinespace[6pt]
    CluSL (Section~\ref{sec:clusl}) & 
    The number of clusters is fixed. Cluster assignments are estimated deterministically while taking dissimilarity between covariate profiles into account. References:  
    \cite{chenClusterAwareSupervisedLearning2022}. \\
    \midrule 
    \textbf{Agnostic-cluster} &
    Two-step procedures: In step 1 units are clustered based on their covariate vectors. 
    In step 2 variables derived from the clustering in step 1 are used (e.g., as covariates) in the model for the outcome. Clustering-derived variables can be cluster membership indicators, 
    dissimilarities to cluster centroids, or cluster membership probabilities. \\
    \addlinespace[6pt]
    Model-based (Section~\ref{sec:outcome_agnostic}) & 
    The number of clusters is fixed. Cluster assignments in step 1 are based on a finite-mixture model for the covariates. References: 
    \cite{bayerNetworkbasedClusteringUnveils2025, esmailiMultichannelMixtureModels2021, gronsbellAutomatedFeatureSelection2019, ramachandranAssessingValueUnsupervised2021}. \\ 
    \addlinespace[6pt]
    Algorithmic (Section~\ref{sec:outcome_agnostic}) & 
    The number of clusters is fixed. Cluster assignments in step 1 are based on algorithmic clustering methods (e.g., $k$-means, hierarchical clustering). References:
    \cite{alexanderUsingTimeSeries2018, almeidaLikelyLightAccurate2023, canahuateSpatiallyawareClusteringImproves2023a, choMetabolicPhenotypingComputed2024a, dingEvaluatingPrognosticValue2024, eberhardDeepSurvivalAnalysis2024, gaoFeatureReductionText2012a, gartnerHowPredictiveFuture2024, kellerCovariateadaptiveClusteringExposures2017, liClusterbasedBaggingConstrained2019, liExploitingComplexNetworkBased2024, maIntegratingGenomicSignatures2018, mishraNovelVersionHorse2025, mokhtarAdaptiveBasedMachine2025a, nasiriBoostingAutomatedSleep2021, nguyenMultivariateLongitudinalData2023, palomino-echeverriaRobustClusteringStrategy2024, piernikStudyUsingData2021, poojamrHybridDecisionSupport2015, songMultifeatureClusteringStep2024a, trivediUtilityClusteringPrediction, xuSubgroupBasedAdaptiveSUBA2016}. \\
    \bottomrule
  \end{tabular}
\end{table}

Within the class of agnostic-cluster models
we further distinguish between those that employ model-based clustering and those that employ algorithmic clustering to the covariates in the first step. 
Model-based clustering is based on a probability distribution for the covariates and estimates cluster membership using the likelihood or posterior distribution. Algorithmic clustering groups observations by optimizing a clustering criterion to the covariates, without the specification of an explicit distribution for the covariate distribution. The clustering criterion is typically based on distance- or similarity measures between observational units, 
often motivated heuristically \cite{fraleyModelBasedClusteringDiscriminant2002}. Typical examples of model-based clustering methods are finite mixture models \cite{fraleyModelBasedClusteringDiscriminant2002}. 
Employed in an agnostic-clustering model a finite mixture distribution is  
assumed for the covariates in the first step of the procedure. 
Examples of algorithmic clustering methods are hierarchical clustering, spectral clustering \cite{vonluxburgTutorialSpectralClustering2007} 
and $k$-means clustering.

We also extracted the minima and maxima of both the sample sizes and the number of covariates used for 
clustering from both simulation studies and applications to real data to illustrate typical settings 
in which clustering-based outcome models are applied. 

Additional characteristics extracted from each record are listed in 
Table~\ref{tbl:all_data}. These include the objective pursued by employing a 
clustering-based outcome model (\emph{Objective}), the method used to 
determine the number of clusters (\emph{Cluster Number Method}), the 
primary outcome type (\emph{Outcome Type}), whether software code was 
provided and the software language used 
(\emph{Software Language}), and the research field from which data 
examples were drawn, such as biomedical sciences, engineering, or 
computer science (\emph{Data Application Domain}).

\begin{table}
  \centering
  \caption{
  Distribution of categorical characteristics from 55 records included in the scoping review. 
  Category counts exceed 55 because individual records could be assigned to multiple categories within a characteristic.
 \textit{Objective} refers to the purpose of incorporating clustering into the model; 
  \textit{Cluster number selection} indicates the approach used to determine or infer the number of clusters; \textit{Outcome Type} describes the nature of the outcome variable for which the model was developed; 
  and \textit{Software Language} count the number of records that reported the use of software and the specific included software language used, respectively. \emph{Data Application Domain}
  represents the research field from which data were used to illustrate the method.
  }
  \label{tbl:all_data}
    \begin{tabular}{rlc}
      \toprule
      Characteristic & Category & Count \\ 
      \midrule
Objective 
    & Subgroup identification & 31 \\ 
& Dimensionality reduction & 21 \\ 
& Feature extraction & 20 \\ 
& Addressing heterogeneity & 9 \\ 
& Improve prediction & 6 \\ 
& Include longitudinal covariates & 3 \\ 
& Account for missing data & 1 \\ 
& Variable selection & 1 \\ 
Cluster number selection
& Posterior distribution & 20 \\ 
& Fixed & 11 \\ 
& BIC & 7 \\ 
& AIC & 5 \\ 
& Cross validation & 5 \\ 
& Gap statistic & 3 \\ 
& Silhouette & 3 \\ 
& Automatic & 2 \\ 
& Elbow method & 2 \\ 
Outcome type 
& Any & 18 \\ 
& Metric & 14 \\ 
& Time-to-event & 12 \\ 
& Categorical & 10 \\ 
Software language & R & 15 \\ 
& Python & 3 \\ 
& Julia & 1 \\ 
& Other & 3 \\ 
& No software code included & 38\\
Data application domain 
& Biomedical sciences & 36 \\ 
& Finance/economics & 7 \\ 
& Engineering & 5 \\ 
& Public health & 5 \\ 
& Environmental & 4 \\ 
& Behavioural/social sciences & 3 \\ 
& Natural language processing & 1 \\ 
& Other & 4 \\
      \bottomrule
  \end{tabular}
\end{table}

\clearpage
\newpage

\section{Results} 
\label{sec:results}

Following the screening process, 55 of the 98 fully screened records were selected for inclusion (see Table~\ref{tbl:studies} in the Appendix for 
a list of all included records, and supplement S4 for a CSV file of all records and their coded characteristics). Initially, human reviewers recommended only 50 records for inclusion, 
whereas Elicit recommended 82 (Table~\ref{tbl:agreement}). 

\begin{table}
  \caption{\enspace 
  Agreement between AI-assisted (Elicit) and human screening decisions for inclusion of records in the review. 
}
  \label{tbl:agreement}
  \centering
  \begin{tabular}[t]{lrrrr}
  \toprule
  \multicolumn{1}{c}{} & \multicolumn{4}{c}{Human Reviewers} \\
  \multicolumn{1}{c}{} & \multicolumn{2}{c}{Initial Decisions} & \multicolumn{2}{c}{Decisions After Revision} \\
  \cmidrule(l{3pt}r{3pt}){2-3}  \cmidrule(l{3pt}r{3pt}){4-5} 
  Elicit & Exclude & Include & Exclude & Include\\
  \midrule
         Exclude & 16 & 0  &  16 & 0\\
         Include & 32 & 50 &  27 & 55\\
  \bottomrule
  \end{tabular}
\end{table}

Following discussion among selected reviewers, five of the 32 discordant records had their human decision 
overturned, yielding the final corpus. 
The remaining 27 discordant decisions were 
all falsely recommended for inclusion (no records were falsely recommended for exclusion). 
Elicit seemingly applied an "any criterion met" logic rather than requiring all 
inclusion criteria to be satisfied. This observation is contrary to other reports 
of LLMs being too stringent leading to an influx of falsely recommending the 
exclusion of records \cite{delgado-chaves2025a}.

The 27 discordant records were excluded for the following reasons: 
mentioning clustering only, without any reference to a model for an outcome (12), 
proposing methods for clustering covariates rather than observational units (4), 
not mentioning clustering at all or using the term "clustering" to mean something different than a 
data-driven method for grouping observations (e.g., referring to a-priori defined subgroups as clusters, 4), 
lacking sufficient description of the method(s) (5), 
purely applied works (1), reviewing existing methods (1). As a measure of agreement between Elicit and 
human reviewers Cohen's\cite{cohen1960} $\kappa$ was 0.34 for initial decisions and improved to 0.40 after 
the revision of discordant decisions.

\begin{figure}[H]
  \centering
  \includegraphics{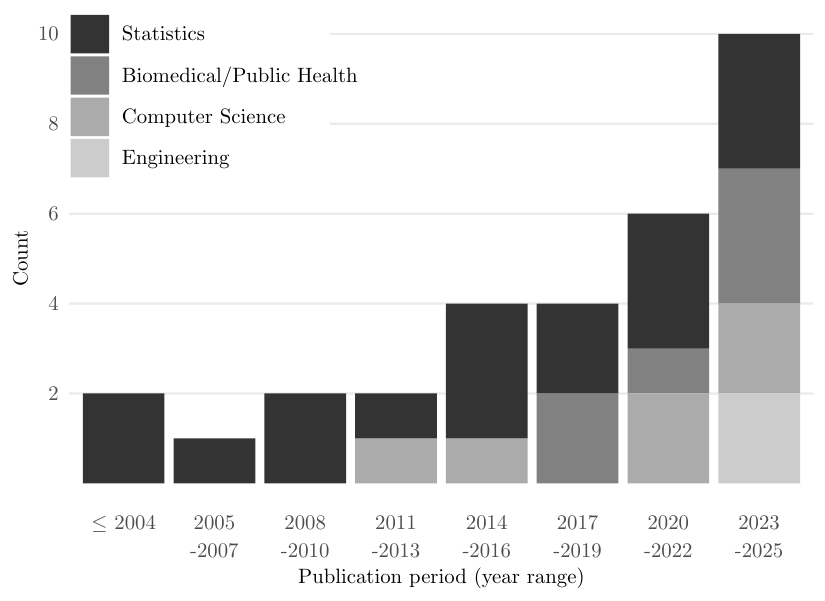}
\caption{Number of records by year range (period) and discipline of publication sources (49 journal articles, 5 conference papers, and 1 preprint).}
\label{fig:publication_counts} 
\end{figure}

Five of the 55 records we included were published before 2011, all in statistical journals (Figure~\ref{fig:publication_counts}). 
From 2011 onward, publication sources other than  statistical ones begin to appear as outlets, particularly 
computer science journals and computer science conferences as well as biomedical science journals. 
Figure~\ref{fig:classification}  shows the number of records that included a given clustering-based outcome model by whether it is an informed-cluster or agnostic-cluster model and - for agnostic-cluster models - whether model-based or 
algorithmic clustering was employed in the first step. Agnostic-cluster models were slightly more prevalent than informed-cluster models. Three groups of informed-cluster models were identified: 
(1) \textit{product-partition models} (PPM) \cite{barry1992} and \textit{product-partition models with covariates} (PPMx) \cite{mullerProductPartitionModel2011}, 
(2) \textit{finite mixtures of regression models} (FMR)  \cite{leischFlexMixGeneralFramework2004}, 
and (3) \textit{cluster-aware supervised learning} (CluSL) \cite{chenClusterAwareSupervisedLearning2022}. Within agnostic-cluster
models, only four employed model-based clustering (all finite-mixture models) in the first step of the 
analysis. Among the remaining algorithmic clustering methods, $k$-means was most prevalent, followed by hierarchical 
clustering, $k$-medoids and spectral clustering. The category "Other" Figure~\ref{fig:classification} 
comprises cosine-similarity clustering \cite{gaoFeatureReductionText2012a}, fuzzy $C$-means clustering and subtractive clustering \cite{poojamrHybridDecisionSupport2015}, and a three-step clustering approach for 
longitudinal data \cite{nguyenMultivariateLongitudinalData2023} originally proposed by Leffondre et al. \citep{leffondreStatisticalMeasuresWere2004}. 
In Section~\ref{sec:key_approaches} we give a  detailed description of key clustering-based outcome models identified.

\begin{figure}[H]
  \centering
  \includegraphics{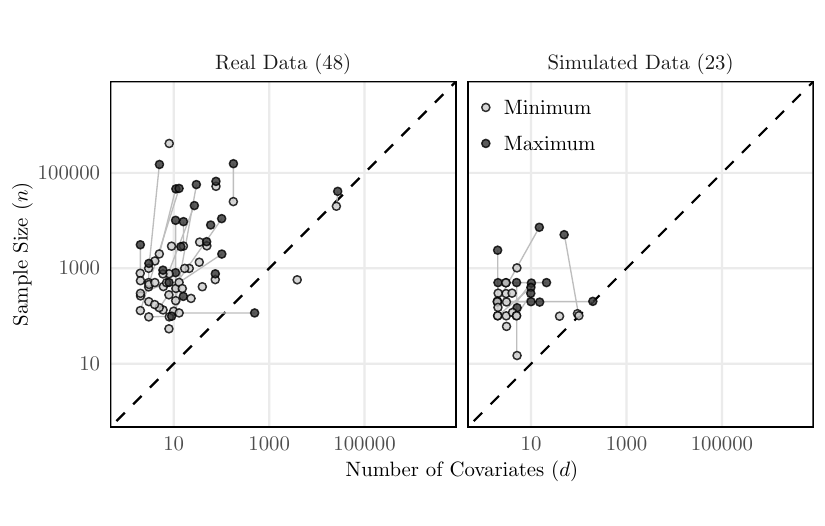}
  \caption{
    Scatter plot of sample sizes ($n$) against number of covariates ($d$) across records using real data (48 records, left) and simulated data (23 records, right). Each record contributes up to two data points: one for the minimum number of covariates paired with the minimum sample size (light grey), and one for the maximum number of covariates
    paired with the maximum sample size (dark grey). The solid grey lines connect data points from the same study. The dashed line represents $n = d$. Both axes use $\log_{10}$ scales.}
\label{fig:scatter}
\end{figure}

As demonstrated in Figure~\ref{fig:scatter}, the real data exhibited larger sample sizes and greater numbers of covariates in comparison to the simulated data. A total of 48 records  
contained applications of clustering-based outcome models to real-world data, while 23 employed simulations. 
The real-world data sets under consideration even comprise examples where the number of covariates exceeds the sample size ($n < d$), thereby underscoring 
the notion that clustering can be utilised as a tool for analysing these types of high-dimensional data by 
reducing the dimensionality of the covariates. 
The simulation studies encompass a wide range of scenarios, including settings with a sample size of less than 100, as well as scenarios where the sample size and the number of covariates 
are similar.
Note that the categories within each characteristic in Table~\ref{tbl:all_data} are not mutually exclusive - a given record can fall into multiple categories per characteristic. 

Among the extracted goals of clustering-based outcome models, subgroup identification was mentioned most often, followed by 
dimensionality reduction and feature extraction. In our coding, subgroup identification refers to using the clusters obtained 
as candidates for prognostic subgroups; that is, individuals within clusters show more similar outcomes than individuals across clusters.
Agnostic-cluster models provide a particularly intuitive approach to achieving this goal: Covariate information is used to group patients into clusters based on covariate similarity. The resulting clusters are then treated as candidates for subgroups and cluster-specific models are fitted for the outcome (see Section~\ref{sec:applications} for examples of this in the biomedical sciences).

Addressing heterogeneity is a related, or overarching goal. Forming 
subgroups represents one popular approach for explaining heterogeneity in the outcome (or in treatment responses), 
but some clustering-based outcome models also address heterogeneity more subtly. In particular, PPMx 
(see Section~\ref{sec:key_approaches} for a mathematical description) captures heterogeneity by modelling the number of 
clusters as random and assuming cluster-specific sampling or regression models for the outcome. PPMx can 
also be employed to select subsets of the available covariates \cite{barcellaVariableSelectionCovariate2016},
a goal which we refer to as 'variable selection' in Table~\ref{tbl:all_data}. The goal of dimensionality reduction refers to the idea that using cluster assignments or functions of cluster assignments obtained from high-dimensional covariate 
vectors effectively compresses information contained in these covariates to a smaller 
number of covariates. 
Feature extraction refers to the derivation of new features or covariates based on clustering, which 
can be used for dimensionality reduction but also for adding covariates to the model for the outcome. An 
example of the latter can be found in \cite{piernikStudyUsingData2021} where existing covariate 
vectors are augmented to include dissimilarities between covariate vectors and cluster centroids (see also the paragraph on agnostic-cluster models in Section~\ref{sec:key_approaches}). Another example of feature extraction is
the clustering of units based on longitudinal covariates, which is also mentioned as a separate goal in Table~\ref{tbl:all_data}. 
Here, cluster membership indicators can be regarded as extracted features (in this case, subgroup candidates) to be used as covariates in the 
model for the outcome. One record \cite{pageClusteringPredictionVariable2022}
highlighted the use of clustering as a method to account for missing covariate values. 

Among methods for inferring the number of clusters, the category 
\emph{Posterior Distribution} occurred most often. This category subsumes 
methods that yield a posterior distribution for the number of clusters 
and primarily concerns records employing PPMx. \emph{Fixed} denotes cases 
in which the number of clusters was set to a pre-specified value. 
\emph{BIC}, \emph{AIC}, and \emph{Cross Validation} include cases where 
the number of clusters was selected from a range of candidate values 
using the Bayesian information criterion, Akaike's information criterion, 
or cross-validation, respectively. Other methods include automatic 
approaches (e.g., affinity propagation), the elbow method, and the 
silhouette method.

The types of outcomes for which the clustering-based outcome models have been developed or applied include \emph{metric} 
(not necessarily continuous), \emph{time-to-event}, and \emph{categorical}. 
The fourth category, denoted by \emph{Any}, indicates models that did not mention any constraints regarding the outcome type. 
Data examples from the biomedical sciences clearly outnumbered those from all other domains. 

Informed-cluster models were published almost exclusively in statistical journals (see Table~\ref{tbl:by_discipline}) with only one informed-cluster model \citep{chenClusterAwareSupervisedLearning2022} published in a journal labelled as belonging to the discipline of computer science. Supplementary materials S5 contains additional tables for objectives and methods used to determine the number of clusters. 
\begin{table}[!h]
  \centering
  \caption{
    Number of informed-cluster and agnostic-cluster models by discipline of publication source. 
  }
  \label{tbl:by_discipline}
\begin{tabular}{rcccc}
\toprule
  Model & Biomedical/Public Health & Computer Science & Engineering & Statistics\\
\midrule
    Agnostic-cluster & 13 & 10 & 3 & 6\\
    Informed-cluster & 0 & 1 & 0 & 27\\
\bottomrule
\end{tabular}
\end{table}
\newpage 
\section{Key Approaches}\label{sec:key_approaches}
In this section we describe the identified clustering-based outcome models in more detail. As above, we distinguish between informed-cluster and agnostic-cluster models, as well as between model-based and algorithmic clustering. 

Let $[n] := \{1,\ldots,n\}$ denote the index set of observational units. Given a fixed number $k$, a 
\emph{$k$-partition} or \emph{$k$-clustering}, $\rho_{n,k}$, of $[n]$ is defined as a collection of nonempty, disjoint and exhaustive subsets 
of $[n]$; that is, $\rho_{n,k} = \{S_1,\ldots,S_k\}, S_j\subseteq [n]$ such that $S_j\cap S_{\ell} = \emptyset$, $j\neq \ell$, $\bigcup_{j = 1}^k S_j = [n]$. 
The individual sets $S_1,\ldots,S_k$ in the $k$-partition are referred to as \emph{clusters}. The set of all possible partitions of $[n]$ 
is denoted by $\mathcal{P}_n$. The subscript $k$ will be omitted when referring to a partition of arbitrary size, in which case 
we write $\rho_n \in \mathcal{P}_n$. 
For a given partition, $\rho_{n,k} = \{S_1,\ldots,S_k\}$, for each $i$ define 
$\boldsymbol{s}_i$ to be the $k$-dimensional vector with $s_{ij} = 1$ if unit $i$
is in $S_j$ and $0$ otherwise.
This way, any $k$-partition $\rho_{n,k}\in \mathcal{P}_n$ can be 
represented by a collection of $n$ vectors $\boldsymbol{s}_1,\ldots, \boldsymbol{s}_n$ each of dimension $k$. 

The covariate vector for unit $i$ is denoted by $\boldsymbol{x}_i = (x_{i1},\ldots, x_{id})$, where $d$ denotes 
the number of covariates (which is assumed to be the same for all units). The outcome models under consideration  
incorporate clusterings of $[n]$ based on $\boldsymbol{x}_i$ or subsets of $\boldsymbol{x}_i$ for an outcome denoted 
by $Y_i$, which need not be metric and can be a vector itself. 

We assume that the $Y_i$'s follow a distribution conditional on the covariates $\boldsymbol{x}$ - referred to as 
the \emph{sampling model} -  with density $p(y_i|\theta, \boldsymbol{x}_i)$ and arbitrary parameter $\theta$. 
In most cases $p(y_i| \theta, \boldsymbol{x}_i)$ will be a regression model of $Y_i$ on $\boldsymbol{x}_i$.
Densities of distributions of random variables will be denoted by $p$. The sets 
$\boldsymbol{y}_j^* = \{y_i : i\in S_j\}$, $\boldsymbol{x}_j^* = \{\boldsymbol{x}_i: i\in S_j\}$, denote the 
realisations of outcomes and covariates, respectively, of all units in cluster $S_j$. From now on, any cluster-specific 
characteristics are equipped with an asterisk; for instance, parameters specific to cluster $S_j$ are denoted 
by $\theta_j^*$. The collection of all cluster-specific parameters is denoted by 
$\boldsymbol{\theta}^* = \{\theta_1^*, \ldots, \theta_k^*\}$.

In cases where the original presentation of the clustering-based outcome model specifically references a 
\emph{prediction function} from a space $\mathcal{X}$ (usually the covariate space or an augmentation of it) 
to the space $\mathcal{Y}$, where the outcomes $Y_1,\ldots, Y_n$ are realised, we will denote 
this function by $f$, with estimators being denoted by $\hat f$.

\subsection{Informed-cluster models}

The class of informed-cluster models encompasses PPM/PPMx \cite{barry1992, mullerProductPartitionModel2011}, 
FMR \cite{leischFlexMixGeneralFramework2004, grunFlexMixVersion22008, proust-limaEstimationExtendedMixed2017} and 
CluSL \cite{chenClusterAwareSupervisedLearning2022}. These models 
specify cluster-specific sampling models for the outcome with cluster assignments depending both on the covariates and the outcome.

\subsubsection{Product-Partition Models and Product-Partition Models with Covariates}
\label{sec:PPMx}

\paragraph{The PPM/PPMx framework.} 

We consider PPMx first and note that a PPM is just a special case of a PPMx. Let $Y^n = (Y_1,\ldots,Y_n)$
denote the vector of all observed outcomes, $y^n$ the corresponding vector of realisations and $\boldsymbol{x}^n = \{\boldsymbol{x}_1,\ldots,\boldsymbol{x}_n\}$
the collection of all $n$ covariate vectors. A PPMx for $Y^n$ may be written using the following hierarchical form:
\begin{equation}
  \begin{split}
\label{eq:ppmx} 
p(y^n| \boldsymbol{\theta}^*, \boldsymbol{x}^n, \rho_n) &= \prod_{j=1}^k \prod_{i\in S_j} p_j(y_i|\theta_j^*, \boldsymbol{x}_i) \\
    \theta_1^*,\ldots,\theta_{k}^*|\rho_{n} &\overset{\text{i.i.d.}}{\sim} G_{\phi} \\ 
    \phi &\sim H_{\psi}\\
  \mathbb{P}(\rho_{n} = \{S_1,\ldots,S_{k}\}|\boldsymbol{x}^n) &\propto \prod_{j=1}^{k} c(S_j) g(\boldsymbol{x}^*_j),\quad \rho_{n} \in \mathcal{P}_n.
  \end{split}
\end{equation}

Beginning in the hierarchy with the definition of the prior distribution $\mathbb{P}$ on the set of all possible partitions $\mathcal{P}_n$ of $[n]$,
a given partition $\rho_{n} = \{S_1,\ldots,S_k\}$ has probability proportional to the product of two components: 
the values of the cohesion function $c(S_j) \geq 0$ that measures how tightly elements within cluster $S_j$ should be grouped, and the values of the similarity function $g(\boldsymbol{x}_j^*)\geq 0$
which quantifies how 'close' the units in cluster $S_j$ are believed to be. For the choice of $c$ and $g$ several options have been proposed \cite{pageCalibratingCovariateInformed2018}. 
A popular cohesion function is given by 
\begin{equation}
  \label{eq:dp_prior}
c(S_j) = M\times (|S_j|-1)!,
\end{equation}
where $M>0$ is a user-specified concentration parameter and $|S_j|$ denotes the cardinality of the set $S_j$.
While other cohesion functions are possible, the popularity of this particular form 
is due to the correspondence between the prior implied by the cohesion function in Equation~\eqref{eq:dp_prior} and 
a Dirichlet process prior on the parameters of an infinite mixture model for $Y_1,\ldots, Y_n$
when used in a PPM (that is, $g(\boldsymbol{x}_j^*) = 1$, see e.g., Barcella et al. \citep{barcella2017} and Quintana and Iglesias \citep{quintanaBayesianClusteringProduct2003} for further details)

The prior in Equation~\eqref{eq:ppmx} operationalises the intuitive notion that units with similar covariate values should be grouped together by assigning them higher prior probabilities of co-clustering as opposed to units with more dissimilar covariate vectors. At 
the second stage - conditional on a particular $k$-partition -- cluster-specific parameters $\theta_1^*,\ldots,\theta_k^*$ are drawn from a distribution $G_{\phi}$, indexed by a parameter $\phi$. For example $G_{\phi}$ could denote the distribution function 
of a normal distribution with parameter $\phi = (\mu, \sigma^2)$. The parameter $\phi$ is itself equipped 
with a prior distribution $H_{\psi}$ with hyperparameter $\psi$. The top line in Equation~\eqref{eq:ppmx} defines the sampling model for the outcomes
$Y^n$. Conditional on the covariates and the $k$-partition $\rho_{n,k}$ the $Y_i$'s are generated according to cluster-specific probability models $p_1,\ldots,p_k$, indexed by 
the parameters $\theta_1^*,\ldots,\theta_k^*$. 
Equation~\eqref{eq:ppmx} shows that PPMx can include covariate information through both, the prior over $\mathcal{P}_n$ and the sampling model.
A PPM is obtained by setting $g(\boldsymbol{x}_j^*) = 1$, in which case the covariates inform the clustering solely through the sampling model.

Predictions for a future observation $Y_{n+1}$ with covariate vector $\boldsymbol{x}_{n+1}$ are obtained by drawing samples
from the posterior predictive distribution $p(y_{n+1}|\boldsymbol{x}_1,\ldots,\boldsymbol{x}_n, \boldsymbol{x}_{n+1}, y_1,\ldots,y_n)$ \cite{mullerProductPartitionModel2011, pageClusteringPredictionVariable2022}.

\paragraph{PPM/PPMx Applications.} Examples of applications of PPM/PPMx comprise modelling of zero-inflated count data 
\cite{barcellaBayesianNonparametricModel2016}, variable selection problems \citep{barcellaVariableSelectionCovariate2016}, (joint) modelling of time-to-event outcomes 
\citep{martinez-vargasPottsCoxSurvivalRegression2023,leiJointAnalysisTwo2025, mullerProductPartitionModel2011}, convex-mixture regression \citep{lijoiPitmanYorMultinomial2020} and estimation of 
cluster-specific treatment effects for a longitudinal outcome using baseline covariates for the clustering of patients \citep{mullerBayesianInferenceLongitudinal2014}.

For instance, Barcella et al. \cite{barcellaBayesianNonparametricModel2016} proposed a zero-inflated Poisson (ZIP) model 
for white blood cell counts (WBC) to develop a prognostic tool for urinary tract infection based on lower urinary tract 
symptoms. Here, the cluster-specific model for the WBC counts is given by
 
\begin{equation}
  p(\boldsymbol{y}^n| \boldsymbol{\mu}^*, \boldsymbol{\lambda}^*, \rho_n, \boldsymbol{x}_i) = \prod_{j=1}^k \prod_{i\in S_j} \{(1-\mu_j^*)\delta_0(y_i) + \mu_j^* \text{Poisson}(y_i|\lambda_j^*)\},
\end{equation}

where $\text{Poisson}(\cdot|\lambda_j^*)$ denotes the probability function of a Poisson distribution with parameter $\lambda_j^*$, 
$\delta_0(\cdot)$ denotes the Dirac function placing unit mass at $0$ and $\mu_j^*\in [0,1]$ a mixing probability. 

Argiento et al. \cite{argientoClusteringBloodDonors2024} extended PPMx by using a mixture of PPMx priors on $\mathcal{P}$ in 
Equation~\eqref{eq:ppmx} to identify clusters of blood donors. The outcome of interest was the log of the gap time between recurrent donations. In this setting, each blood donor $i$ contributes up to $r_i$ data points, with 
$Y_{it}$ denoting the log gap time between the donation at time $t$ and that of the previous time $t-1$. The sampling 
model for the log gap time of unit $i$ in cluster $S_j$ at time $t$ was given by 
\begin{equation}
  \label{eq:blood_donors}
\begin{split}
  p_j(y_{it}|\beta_0, \beta_t, \alpha_j^*, \xi_j^*, \sigma^{2*}_j, \eta_{i,t}) &= \text{Normal}(\alpha_j^* + \boldsymbol{x}_i^\top \beta_0 + \boldsymbol{x}_{i,t}^\top \beta_t + \xi_j^* \eta_{i,t}, \sigma_j^{2*})\\ 
  \eta_{i,t} &\overset{\text{i.i.d.}}{\sim} \text{Half-Normal}(0,1),
\end{split}
\end{equation}
where $\boldsymbol{x}_{i,t}$ and $\eta_{i,t}$ denote the covariate vector and a latent variable for subject $i$ at time $t$. 

The model in Equation~\eqref{eq:blood_donors} assumes three cluster-specific parameters - an intercept $\alpha_j^*$, 
the coefficient $\xi_j^*$ for the latent Half-Normal variable at time $t$, and the variance $\sigma_j^{2*}$. 
The remaining parameter vectors $\beta_0$ for those covariates $\boldsymbol{x}_i$ that remain constant over time
and $\beta_t$ for time-dependent covariates are assumed to be the same across clusters. The latent variable $\eta_{i,t}$ was added  
to accommodate the skewed distribution of log gap times observed in the data. 

Standard PPMs (i.e., PPMx with $g(\boldsymbol{x}_j^*) = 1$) have been employed by incorporating covariates 
in the sampling model for the outcome only \citep{kyungBayesianAnalysisRandom2017}. 
This is nevertheless an informed-cluster model that uses covariate information, since the sampling model contributes to the clustering (PPMx and PPM induce an outcome-informed clustering solution).

\subsubsection{Finite Mixtures of Regression Models}
\label{sec:FMR}

FMR models form a substantial body of literature with many applications and several comprehensive reviews 
and tutorials available \citep{leischFlexMixGeneralFramework2004, grunFlexMixVersion22008, proust-limaEstimationExtendedMixed2017, proust-limaJointLatentClass2014}.
We outline the basic framework and focus on extensions to cases that permit the inclusion of functional covariates 
in the clustering as well as approaches that model longitudinal and time-to-event outcomes jointly. 
The number of clusters (often referred to as latent classes in the FMR literature) is typically fixed at a pre-specified value $k$; that is, 
only the subset of $k$-partitions, $\mathcal{P}_{n,k}$, of the set of all partitions $\mathcal{P}_n$ is considered. 

\paragraph{Basic FMR models.} 
Every $k$-partition $\rho_{n,k}\in \mathcal{P}_{n,k}$ corresponds to a collection $\{\boldsymbol{s}_1,\ldots,\boldsymbol{s}_n\}$ of $k$-dimensional cluster membership indicator vectors. Using this equivalence, an FMR model can be expressed using the following hierarchical form: 
\begin{equation}
  \label{eq:fmr}
  \begin{split}
    p(y_i| \boldsymbol{\theta}^*,\boldsymbol{s}_i, \boldsymbol{x}_i) &= \prod_{j=1}^k \{p_j(y_i|\boldsymbol{x}_i, \theta_j^*)\}^{s_{ij}}\\
   \boldsymbol{s_i}|\boldsymbol{x_i}, \boldsymbol{\alpha}^* &\overset{\text{i.i.d.}}{\sim} \text{Multinomial}(1|\boldsymbol{\pi}(\boldsymbol{x_i},\boldsymbol{\alpha}^*)), \forall i \in [n],
  \end{split}
\end{equation}

where $s_{ij}\in \{0,1\}$ are cluster-membership indicators and  
$\boldsymbol{\pi}(\boldsymbol{x}_i, \boldsymbol{\alpha}^*) = (\pi_1(\boldsymbol{x}_i, \boldsymbol{\alpha}^*),\ldots,\pi_k(\boldsymbol{x}_i, \boldsymbol{\alpha}^*))$ is 
a vector of cluster probabilities (i.e., $0\leq \pi_j(\boldsymbol{x}_i, \boldsymbol{\alpha}^*) \leq 1$ and $\sum_{j=1}^k \pi_j(\boldsymbol{x}_i, \boldsymbol{\alpha}^*) = 1, \forall i \in [n]$). 
These probabilities depend on the covariates $\boldsymbol{x}_i$ through parameters $\boldsymbol{\alpha^*} = (\alpha_1^*,\ldots,\alpha_{k-1}^*)$, where 
the $\alpha_j^*$'s may be vector-valued. A common choice is a multinomial logistic regression model: 
\begin{equation}
  \label{eq:multinomial}
  \pi_j(\boldsymbol{x}_i, \boldsymbol{\alpha}^*) = \frac{\exp(\alpha_j^{*\top} \boldsymbol{x}_i)}{1 + \sum_{\ell=1}^{k-1}\exp(\alpha_\ell^{*\top} \boldsymbol{x}_i)}, \quad j \in \{1,\ldots,k-1\},
\end{equation}

The formulation in Equation~\eqref{eq:fmr} makes explicit a key difference between PPM/PPMx 
and FMR models: the iid Multinomial distribution on $\boldsymbol{s}_1,\ldots, \boldsymbol{s}_n$ 
induces a distribution over the subset of all partitions of size $k$ which we denote by $\mathcal{P}_{n,k}$. PPMx 
assumes a prior distribution over $\mathcal{P}_n$; that is, all possible partitions of all sizes.

Marginalizing out the cluster indicators yields the finite mixture representation for $Y_i$
\begin{equation}
  \label{eq:fmr_marginal}
  \begin{split}
  p(y_i| \boldsymbol{\alpha}^*, \boldsymbol{\theta}^*, \boldsymbol{x}_i) = \sum_{j=1}^k \pi_j(\boldsymbol{x}_i, \boldsymbol{\alpha}^*) p_j(y_i| \boldsymbol{x}_i, \theta_j^*),
\end{split}
\end{equation}
highlighting the standard finite mixture structure of an FMR. 

We note that three \citep{garciaUnsupervisedBayesianClassification2024, wangMixturesT2024, zhangJointLatentClass2022} of the 
identified articles belonging to the FMR category were part of the original search results. Five additional
records were added manually,  as they are foundational works for these three records. 

\paragraph{FMR models using functional covariates.} Garcia et al. \cite{garciaUnsupervisedBayesianClassification2024} proposed an FMR model that 
accommodates functional covariates, particularly covariates measured over time. In their  
approach, in addition to the cross-sectional covariate vector $\boldsymbol{x}_i$, 
a functional covariate $u_{i}(t)$ defined on a closed support $\tau$ at times $t = t_{i1}, \ldots, t_{ir_i}$ is observed, 
where $r_i$ denotes the total number of measurements on the functional covariate for unit $i$. 
Importantly, the number of measurements need not be the same across units. The $r_i$ observed values from 
$u_i(t)$ are collected in a vector $\boldsymbol{u}_i = (u_{i}(t_{i1}),\ldots, u_{i}(t_{ir_i}))$. Extending this 
framework to multiple functional covariates, suppose there are $m$ functional covariates $u_{i1}(t),\ldots, u_{im}(t)$ defined on closed supports $\tau_1,\ldots,\tau_m$ 
yielding $m$ vectors $\boldsymbol{u}_i^{(1)}, \ldots, \boldsymbol{u}_i^{(m)}$ of lengths 
$r_{i1}, \ldots, r_{im}$, respectively.
Garcia et al. model the cluster probabilities via (treating cluster $k$ as the reference cluster)
\begin{equation}
  \label{eq:functional_covars}
  g\{\pi_{j}(\boldsymbol{x}_i, \boldsymbol{u}_i^{(1)},\ldots, \boldsymbol{u}_i^{(m)}, \boldsymbol{\alpha}^{*})\} =  \alpha_j^{*\top} \boldsymbol{x}_i + \sum_{\ell = 1}^m \int_{\tau_\ell} J_{\ell j}(u_{i\ell}(t), t; \phi_{\ell j}) dt, \quad j \in \{1,\ldots, k-1\},
\end{equation}
where $g$ is a link function, $\alpha_j^*$ contains the regression coefficients for the cross-sectional covariates in cluster $S_j$, and $J_{\ell j}$ is smooth in both arguments $u_{i\ell}$ and $t$ parameterised by $\phi_{\ell j}$. One of 
the natural choices for $J_{\ell j}$, used by Garcia et al., is given by  
\begin{equation}
  \label{eq:linear_form}
J_{\ell j}(u_{i\ell}(t),t;  \phi_{\ell j}) = b_{\ell j}(t; \phi_{\ell j}) u_{i \ell}(t) 
\end{equation}
where the weight function  $b_{\ell j}(\cdot; \phi_{\ell j})$ is expressed as a linear combination of cubic $B$-spline basis functions with 
coefficient vector $\phi_{\ell j}$. 

Garcia et al. applied this model to identify placebo responders in a clinical trial of sertraline, 
a drug targeting major depressive disorder. Placebo responders were defined as patients whose Hamilton Depression Rating Scale (HAM-D) scores improved within one week of treatment initiation. Since the active drug (sertraline) is believed to require at least two weeks to take effect, 
early improvement is likely attributable to placebo effects or spontaneous recovery rather than the drug itself.
Electroencephalography (EEG) measurements taken at 14 electrodes located above different brain 
regions were used as functional covariates to inform the clustering in Equation~\eqref{eq:functional_covars}. 
Here, the functional covariates are indexed by frequency (4-15 Hz) rather than time.

\paragraph{FMR joint models.} FMR-based models are also applicable in the context of joint models for  
time-to-event outcomes and biomarker trajectories, where they are referred to 
as \textit{joint latent class models} \cite{proust-limaJointLatentClass2014, proust-limaDevelopmentValidationDynamic2009}.
Here, a longitudinal outcome $Y_i(t)$, such as a biomarker sequence, measured at times $t = t_1,\ldots, t_{r_i}$  
is connected to a (delayed) time to event variable $W_i$ by modelling the measurements 
$Y_{i1},\ldots,Y_{ir_i}$ of $Y_i(t)$ using an FMR as in Equation~\eqref{eq:fmr_marginal} and assuming cluster-specific baseline-hazard functions in 
the proportional hazards model for the time to event $W_i$. In more detail, let $\boldsymbol{Y}_i = (Y_{i1},\ldots, Y_{ir_i})$ denote 
the vector of measurements on the longitudinal outcome, $e_i$ the event indicator for unit $i$ ($e_i = 1$ if the event occurs, $e_i = 0$ otherwise).
The proportional hazards model for $W_i$ in cluster $S_j$ is given by the hazard function 
\begin{equation}
  \label{eq:prop_hazard}
  h_j(w_i, \boldsymbol{x}_i; \zeta_j^*, \gamma) = h_{0j}(w_i; \zeta_j^*) \exp(\boldsymbol{x}_i^\top\gamma),
\end{equation}
where $h_{0j}(\cdot; \zeta_j^*)$ is the baseline hazard function with parameter $\zeta_j^*$ 
in cluster $S_j$, and $\gamma$ denotes a vector of regression coefficients. 
The following joint model for $(\boldsymbol{Y}_i, W_i)$ is assumed:
\begin{equation}
  p(\boldsymbol{y}_i, w_i| \boldsymbol{\alpha}^*, \boldsymbol{\theta}^*, \boldsymbol{\zeta}^*, \gamma) = \sum_{j=1}^k \pi_j(\boldsymbol{x}_i, \boldsymbol{\alpha}^*) p_j(\boldsymbol{y}_i| \boldsymbol{x}_i, \theta_j^*) h_j(w_i; \zeta_j^*, \gamma)^{e_i}[1-F_j(w_i|\zeta_j^*, \gamma)], 
\end{equation}
where $F_j(\cdot|\zeta_j^*, \gamma)$  denotes the distribution function corresponding to the hazard 
function in Equation~\eqref{eq:prop_hazard}. 
Proust-Lima et al. \cite{proust-limaDevelopmentValidationDynamic2009} used this FMR-based 
joint model to predict prostate cancer recurrence probabilities based on prostate specific 
antigen trajectories (i.e., the longitudinal outcome) after radiation therapy. 

\subsubsection{Cluster-Aware Supervised Learning \citep{chenClusterAwareSupervisedLearning2022}}
\label{sec:clusl}

The presentation of PPM/PPMx and FMR was entirely in terms of hierarchical probability models, where we 
used densities to identify distributions of the cluster-specific sampling model for unit $i$'s outcome $Y_i$. With CluSL 
we again have cluster-specific sampling models $\{p_j(y_i| \theta_j^*): j \in [k]\}$, however, the assignment 
of individuals to clusters is no longer probabilistic, but deterministic. Conditional on a fixed 
$k$-partition $\rho_{n,k}$, the sampling model for $Y_i$ is exactly the same as the first line in Equation~\eqref{eq:fmr} specifying the FMR, 
which may equivalently be written as 
\begin{equation}
  \label{eq:clus_sl}
  p(y_i|\boldsymbol{\theta}^*, \boldsymbol{s}_i, \boldsymbol{x}_i) = \sum_{j=1}^k s_{ij} p_j(y_i| \theta_j^*, \boldsymbol{x}_i).
\end{equation}
Instead of assuming  $\boldsymbol{s}_i \overset{\text{i.i.d.}}{\sim} \text{Multinomial}(1|\boldsymbol{\pi}(\boldsymbol{x}_i, \boldsymbol{\alpha}^*))$ as 
in Equation~\eqref{eq:fmr}, CluSL seeks to estimate the indicators $\{s_{ij}: i\in [n], j\in [k]\}$ (and hence the partition $\{\boldsymbol{s}_i: i\in [n]\}$) 
along with the model parameters $\boldsymbol{\theta}^*$ directly by solving the minimization problem: 
\begin{eqnarray}
  \label{eq:clusl}
  \begin{aligned}
  \min_{\boldsymbol{\theta}^*, \boldsymbol{s}, \boldsymbol{m}_1^*,\ldots,\boldsymbol{m}_k^*} \sum_{i=1}^n \sum_{j = 1}^k  s_{ij} L_j^*(y_i, \boldsymbol{x}_i, \theta_j^*) + \lambda \sum_{i=1}^n \sum_{j = 1}^k  s_{ij} D^*_j(\boldsymbol{x}_i,\boldsymbol{m}_j^*) \\ 
  \text{subject to}  \sum_{j = 1}^k s_{ij} = 1, \forall i \in [n] \\ 
                     \sum_{i = 1}^n  s_{ij} \geq 1, \forall j \in [k] \\ 
                     s_{ij}\in \{0,1\}, \forall i \in [n], \forall j\in [k].
  \end{aligned}
\end{eqnarray}

Here, $L_j^*$ denotes a cluster-specific loss function that takes $(y_i, \boldsymbol{x}_i, \theta_j^*)$ as 
inputs, the vectors $\{\boldsymbol{m}_j^*\in \mathbb{R}^d: j \in [k]\}$ denote cluster-specific representative 
values in covariate space called \textit{centroids}, the function $D_j^*$ denotes the 
\textit{dissimilarity} between a given unit's covariate vector $\boldsymbol{x}_i$ and the centroid $\boldsymbol{m}_j^*$
of cluster $S_j$, and $\lambda \geq 0$ denotes a tuning parameter that controls the influence of dissimilarity between covariate 
vectors and the corresponding cluster centroids. The constraints ensure that each unit is assigned to exactly one cluster and 
that every cluster contains at least one unit. See Chen and Xie \cite{chenClusterAwareSupervisedLearning2022} for further 
details on how Equation~\eqref{eq:clusl} can be solved.

Note that the term $\sum_{i=1}^n \sum_{j=1}^k s_{ij}D_j^*(\boldsymbol{x}_i, \boldsymbol{m}_j^*)$ 
serves to regularize the cluster assignments $\{s_{ij}: i \in [n], j \in [k]\}$ by incorporating 
information about covariate similarity. When a unit $i$ is assigned to cluster $S_j$ (i.e., $s_{ij} = 1$), the 
penalty contribution for that unit is $D_j^*(\boldsymbol{x}_i, \boldsymbol{m}_j^*)$, which increases as the dissimilarity between 
$\boldsymbol{x}_i$ and the cluster centroid $\boldsymbol{m}_j^*$ grows. The penalty term encourages units with 
similar covariate vectors to be assigned to the same cluster. Thus, the optimization in Equation ~\eqref{eq:clusl} balances 
two objectives: (1) minimizing cluster-specific loss functions and (2) promoting cluster assignments that 
group together units with similar covariate vectors. The tuning parameter $\lambda$ controls the relative 
importance of these two objectives. For $\lambda = 0$, the problem reduces to what 
Chen and Xie \cite{chenClusterAwareSupervisedLearning2022} refer to as \textit{supervised-learning-based clustering}, where 
the estimation of cluster indicators and the parameters of the sampling models are carried out based solely on 
the outcome, without incorporating covariate similarity into the clustering. On the other hand, for $\lambda$ large enough, the solution to Equation~\eqref{eq:clusl} is equivalent to an agnostic-cluster model (see Section~\ref{sec:outcome_agnostic} below), where units are first clustered based on covariate 
values and then cluster-specific sampling models are fitted by minimizing cluster-specific loss functions.

While in principle there are no constraints regarding the loss functions $\{L_j^*: j \in [k]\}$ that can be employed, 
in practice the choice of loss functions should be informed by the assumed cluster-specific regression models.
Chen and Xie \cite{chenClusterAwareSupervisedLearning2022} introduce three exemplary loss functions: the negative log-likelihood of a linear model with normal error terms for metric outcomes, the 
negative log-likelihood of a logistic regression model for binary outcomes, and the multiclass hinge loss for 
categorical outcomes with more than two classes. Possible dissimilarity functions mentioned in \cite{chenClusterAwareSupervisedLearning2022} 
include the \textit{squared Euclidean distance} $D_j^*(\boldsymbol{x}_i, \boldsymbol{m}_j^*) = ||\boldsymbol{x}_i - \boldsymbol{m}_j^*||^2$, the 
\textit{cosine distance} $D_j^*(\boldsymbol{x}_i, \boldsymbol{m}_j^*) = 1 - \boldsymbol{x}_i^\top \boldsymbol{m}_j^*/(||\boldsymbol{x}_i||\,||\boldsymbol{m}_j^*||)$, 
and the \textit{Gaussian kernel distance} $D_j^*(\boldsymbol{x}_i, \boldsymbol{m}_j^*) = 1 - \exp(-||\boldsymbol{x}_i - \boldsymbol{m}_j^*||^2 / (2\sigma^2))$ for $\sigma^2 > 0$.

Once estimates $\{(\hat\theta_j^*, \boldsymbol{\hat m}_j^*: j\in [k]\}$ and $\{\hat s_{ij}: i\in [n], j\in [k]\}$, 
have been obtained, predictions for a new unit with covariate vector $\boldsymbol{x}_{n+1}$ are obtained by the 
following procedure: (i) compute the dissimilarities $\{D_j^*(\boldsymbol{x}_{n+1}, \boldsymbol{\hat{m}}_j^*): j\in [k]\}$ 
between $\boldsymbol{x}_{n+1}$ and each estimated cluster centroid; (ii) assign unit $n+1$ to the cluster $S_{j'}$ for which 
$D_j^*(\boldsymbol{x}_{n+1}, \boldsymbol{\hat{m}}_j^*)$ is smallest by setting the corresponding element of $\hat s_{(n+1)j'}$ 
equal to $1$ (all other $\hat s_{(n+1)j},$ are set to $0$); and (iii) evaluate the prediction function:
\begin{equation}
  \label{eq:clusl_predict}
  \hat{Y}_{n+1} = \hat{f}_{j'}^*(\boldsymbol{x}_{n+1}; \hat{\theta}_{j'}^*).
\end{equation}

For metric outcomes with $\hat{\theta}_j^* = (\hat{\beta}_j^*, \hat{\sigma}_j^{2*})$, the prediction function may take the form of 
a linear model:
\begin{equation}
  \label{eq:predict_metric}
\hat{f}_j^*(\boldsymbol{x}_{n+1}; \hat{\theta}_j^*) = \boldsymbol{x}_{n+1}^\top \hat{\beta}_j^*.
\end{equation}
Note that Equation~\eqref{eq:predict_metric} does not require the estimate $\hat{\sigma}_j^{2*}$, but only $\hat{\beta}_j^*$. There are, 
however, examples of prediction functions that require both parameter estimates $(\hat{\beta}_j^*, \hat{\sigma}_j^{2*})$, including 
confidence intervals for $\mathbb{E}[Y_{n+1}] = \boldsymbol{x}_{n+1}^\top \beta_j^*$ and prediction intervals for $Y_{n+1}$.
The number of clusters $k$ and the tuning parameter $\lambda$ can be chosen by comparing a suitable criterion, such as the 
mean-square error on test data, across different combinations of $k$ and $\lambda$.
Chen and Xie \cite{chenClusterAwareSupervisedLearning2022} applied the CluSL approach to several of the 
University of California, Irvine (UCI) machine learning data sets, covering a range of metric as well as 
categorical outcomes.

\subsection{Agnostic-cluster models} 
\label{sec:outcome_agnostic}

Agnostic-cluster models can also be described as sequential procedures comprising two steps - 
a clustering step and a modelling step. In the first step, observations are clustered using only their covariate vectors 
$\boldsymbol{x}_1, \ldots, \boldsymbol{x}_n$. Both, model-based or algorithmic clustering methods can be employed in this step. 
The second step consists of fitting a model for the outcome $Y_i$ using additional covariates derived from the clustering 
in the first step. 

\paragraph{Agnostic-cluster models using algorithmic clustering.} Trivedi et al. \cite{trivediUtilityClusteringPrediction} proposed a so-called 
\emph{algorithmic ensemble prediction} approach. Let the number of clusters $k$ be fixed and specify 
cluster-specific regression models: 
\begin{equation}
  \label{eq:trivedi}
  \hat Y_i = \sum_{j=1}^k \hat s_{ij}\hat f_j^*(\boldsymbol{x}_i),
\end{equation}
where the cluster assignments $\{\hat s_{ij}: i \in [n], j \in [k]\}$ are estimated in a first step using $k$-means, 
and $\{\hat f_j^*(\cdot): j \in [k]\}$ are prediction functions obtained by fitting, e.g., a linear model
to the observations within each cluster. $K$-means clustering seeks to solve 
\begin{equation}
  \label{eq:k_means}
  \min_{S_1,\ldots, S_k} \sum_{j = 1}^k \sum_{i \in S_j} ||\boldsymbol{x}_i - \boldsymbol{m}_j^*||_2^2, \quad \boldsymbol{m}_j^* = \frac{1}{|S_j|} \sum_{i\in S_j} \boldsymbol{x}_i,
\end{equation}
where $||\cdot||_2$ denotes the Euclidean norm and the cluster centroid $\boldsymbol{m}_j^*$ is the coordinate-wise average of 
covariate vectors in cluster $S_j$. To obtain a prediction for a new observational unit 
with covariate vector $\boldsymbol{x}_{n+1}$, observation $n+1$ is first assigned to the cluster $S_{j'}$ for which the distance 
$||\boldsymbol{x}_{n+1} - \hat{\boldsymbol{m}}_{j'}^*||_2^2$ between $\boldsymbol{x}_{n+1}$ and the estimated centroid is 
minimal, setting the corresponding cluster indicator $\hat s_{(n+1)j'} = 1$. Plugging $(\hat s_{(n+1)j'}, \boldsymbol{x}_{n+1})$ into 
Equation~\eqref{eq:trivedi} yields the prediction 
\begin{equation}
  \label{eq:predict_trivedi}
  \hat Y_{n+1}^{(k)} = \sum_{j = 1}^k \hat s_{(n+1)j} \hat f_j^*(\boldsymbol{x}_{n+1}) = \hat f_{j'}^*(\boldsymbol{x}_{n+1}).
\end{equation}
This process is repeated for a range of values $k_{\min}\leq k\leq k_{\max}$, and the final prediction for 
unit $n+1$ is given by 
\begin{equation}
  \label{eq:ensemble}
  \hat Y_{n+1} = g\bigg(\hat Y_{n+1}^{(k_{\min})}, \ldots, \hat Y_{n+1}^{(k_{\max})}\bigg), 
\end{equation}
where $g(\cdot)$ is a function that combines the predictions, typically a simple average. Trivedi et al. \cite{trivediUtilityClusteringPrediction}
compared the performance of this approach using 5-fold cross-validation for different prediction functions $\hat f_j^*$ obtained from (1) linear 
regression, (2) stepwise linear regression, and (3) random forests. 
The ensemble predictions given by Equation~\eqref{eq:ensemble} significantly outperformed predictions obtained with $k = 1$ (i.e., no clustering) on most of the 11 datasets from the UCI Machine Learning Repository considered in their study.    

As another example, Piernik and Morzy \cite{piernikStudyUsingData2021} found that feature maps which augment the original covariates by a dissimilarity measure between the original covariate values to each of $k$ cluster centroids can 
improve classification accuracy for a number of classification/discrimination 
methods. That is, letting $\boldsymbol{m}_1^*, \ldots, \boldsymbol{m}_k^*$ denote cluster centroids 
of the $k$ clusters $S_1,\ldots,S_k$ obtained in the first step, the covariate vector for each individual is 
expanded to consist of 
\begin{equation}
  \boldsymbol{\tilde x}_i = (\boldsymbol{x}_i^\top, D(\boldsymbol{x}_i, \boldsymbol{m}_1^*), \ldots,  D(\boldsymbol{x}_i, \boldsymbol{m}_k^*)),
\end{equation}
where $D(\cdot, \cdot)$ again quantifies the dissimilarity between $\boldsymbol{x}_i$ and $\boldsymbol{m}^*_j, j \in [k]$.

Ma et al.\ \citep{maIntegratingGenomicSignatures2018} achieve dimensionality reduction while also avoiding the need 
for cluster-specific regression models, and hence the assumption of cluster-level exchangeability altogether, by 
using a resampling-based clustering approach (consensus clustering) to derive similarity measures between all pairs of observational 
units. These similarity values (between 0 and 1) are then used in the construction of unit-level power priors. In more 
detail, the $n = 116$ patients diagnosed with lung squamous cell carcinoma (LUSC) along with their expressions on $d = 95$ genes (i.e., the covariates) 
were resampled 100 times and the units in each resampled data set were clustered using an algorithmic clustering method 
($k$-means, hierarchical clustering or partitioning around medoids). 

Let $g_{i\ell}\in [0,1]$ denote the proportion of times patients $i$ and $\ell$, $i\neq \ell$, are co-clustered 
across simulations; this $g_{i\ell}$ is a similarity measure between patients $i$ and $\ell$. The collection of  
all pairwise similarities is given by $\{g_{i\ell}: i,\ell \in [n], 
i\neq \ell\}$. Let $W_i$ be the overall survival time of patient $i$ with event indicator $e_i$ (as in Equation~\ref{eq:prop_hazard}), 
let $\boldsymbol{x}_{1:i-1} = \{\boldsymbol{x}_{\ell}: \ell = 1,\ldots, i-1\}$ denote the collection of covariate 
vectors of all patients before patient $i$, and let $A_i \in \{1,2\}$ denote the treatment assigned to patient $i$. 
Assuming a parametric model for $W_i$ with treatment-specific parameter $\theta_a$, the power prior for patient $i$ under 
treatment $A_i = a$ proposed by Ma et al.\ is of the form 
\begin{equation}
  \label{eq:power_prior}
  p_{1:i-1}(\theta_a) \propto \prod_{\ell < i: A_\ell = a} 
  \bigg[\{p(w_{\ell}|\theta_a)\}^{e_\ell} 
  \{1-F(w_\ell|\theta_a)\}^{1-e_\ell} \bigg]^{g_{i\ell}} 
  q(\theta_a), 
\end{equation}
where $q(\theta_a)$ is the density of an initial prior distribution 
for $\theta_a$, and $p(\cdot|\theta_a)$ and $F(\cdot|\theta_a)$ 
denote the density and cumulative distribution function of 
$W_\ell$ under $A_\ell = a$, $\ell \in \{1,\ldots,i-1\}$. 
That is, the power prior for patient $i$ is constructed from the 
likelihood contributions of all previously treated patients under 
treatment $a$, with each contribution weighted by its similarity 
to patient $i$. Thus, previous patients with more similar profiles to patient $i$ receive more weight 
as opposed to patients who are less similar.

Moreover, the predictive probability of patient $i$ under treatment $A_i = a$ exceeding survival time $w_0$ can now be 
computed as 
\begin{equation}
  \label{eq:ppd}
  \mathbb{P}(W_i > w_0, e_i = 0| \boldsymbol{x}_{1:i-1}, \boldsymbol{x}_i, A_i = a) = \int \{1 - F(w_0|\theta_a)\} p_{1:i-1}(\theta_a) d\theta_a.
\end{equation}

This way, heterogeneity is addressed by recommending the treatment with the highest predictive probability 
to patient $i$ and thus implicitly assigning a patient to a predictive subgroup. 
With respect to the dimensionality reduction achieved, Ma et al. used a simple exponential failure time model for the 
survival times, implying that $\theta_a$ (for $a\in \{1,2\}$) was 
just a single scalar parameter. This is a pronounced contrast to the dimensionality of $\theta_a$ that would be required if all (or a 
subset) of the $d = 95$ genes were to be included as covariates.

\paragraph{Agnostic-cluster models using model-based clustering.} Model-based clustering is usually synonymous with assuming a finite mixture model for the covariates; that is 
\begin{equation}
  \label{eq:finite_mixture}
  p(\boldsymbol{x}_i|\boldsymbol{\theta}^*) = \sum_{j=1}^k \pi_j p_j(\boldsymbol{x}_i|\theta_j^*),
\end{equation}
where $\pi_j$ denotes the probability of drawing an observation from 
cluster $S_j$, and $p_j(\cdot|\theta_j^*)$ denotes the model for the covariate vector in cluster $S_j$. 
Ramachandran et al.\ \cite{ramachandranAssessingValueUnsupervised2021} 
used model-based clustering to reduce the dimensionality of candidate covariates for predicting \textit{persistent high users/utilisers} (PHUs) of health care services - a small subset of individuals who account for a 
disproportionate share of health care utilisation. In particular, the authors applied latent class analysis (LCA) to cluster patients 
based on pre-defined groups of diagnostic codes and medication-based morbidity indicators. LCA is a finite mixture model 
for categorical data. For unit $i$ with covariates $\boldsymbol{x}_i$, the finite mixture model in Equation~\eqref{eq:finite_mixture} 
is given by 
\begin{equation}
  \label{eq:lca}
  p(\boldsymbol{x}_i) = \sum_{j=1}^k \pi_j \prod_{\ell=1}^d p_{j\ell}(x_{i\ell}|\theta_{j\ell}^*),
\end{equation}

where $p_{j\ell}(\cdot|\theta_{j\ell}^*)$ is the model for the $\ell$-th component of the covariate vector 
in cluster $S_j$. Note that, the model in Equation~\eqref{eq:lca} assumes that the individual covariate components 
are independent conditional on the cluster membership. This assumption is not required in general: the finite 
mixture model in Equation~\eqref{eq:finite_mixture} specifies cluster-specific joint models for the entire covariate vector.
Moreover, since each covariate $x_{i\ell}$ is categorical, the component densities $p_{j\ell}$ are discrete distributions.
After fitting the model and obtaining estimates $\{\hat \pi_{j}: j \in [k]\}$ and $\{\hat \theta_{j\ell}: j \in [k], \ell\in [d]\}$, estimates 
of posterior class membership probabilities can be obtained using Bayes' theorem:
\begin{equation}
  \label{eq:bayes}
  \hat{p}_{ij} := \widehat{\mathbb{P}}(i \in S_j | \boldsymbol{x}_i) = \frac{\hat{\pi}_j \prod_{\ell=1}^d p_{j\ell}(x_{i\ell}|\hat{\theta}_{j\ell}^*)}{\sum_{m=1}^k \hat{\pi}_m \prod_{\ell=1}^d p_{m\ell}(x_{i\ell}|\hat{\theta}_{m\ell}^*)}.
\end{equation}
The estimated posterior probabilities from Equation~\eqref{eq:bayes} were then used as covariates in a logistic regression model to predict PHU status. This two-stage approach reduced the dimensionality of the original 178 diagnostic code and 
morbidity indicators to a small number of covariates.

\section{Applications in Biomedical Sciences and Public Health}
\label{sec:applications}

In this section we present all applications of clustering-based outcome models found in biomedical and public health
journals (Table~\ref{tbl:impact}). Notably, in all these examples agnostic-cluster models were used.

\begin{longtable}[t]{>{\raggedleft\arraybackslash}p{4cm}>{\raggedright\arraybackslash}p{7cm}>{\raggedright\arraybackslash}m{2cm}}
  \caption{Records included in the scoping review by discipline of publication sources.}
\label{tbl:impact}\\
\toprule
Discipline & Publication Source & Publications \\
\midrule
\endfirsthead
\multicolumn{3}{@{}l}{\textit{(continued)}}\\
\toprule
Discipline & Publication Source & Publications\\
\midrule
\endhead

\endfoot
\bottomrule
\endlastfoot
Statistics & Annals of Applied Statistics & \cite{pageDiscoveringInteractionsUsing2021,kellerCovariateadaptiveClusteringExposures2017}\\
 & Bayesian Analysis & \cite{pagePredictionsBasedClustering2015}\\
 & Biometrics & \cite{argientoClusteringBloodDonors2024,gronsbellAutomatedFeatureSelection2019}\\
 & Biometrika & \cite{lijoiPitmanYorMultinomial2020}\\
 & Biostatistics & \cite{mullerBayesianInferenceLongitudinal2014,proust-limaDevelopmentValidationDynamic2009}\\
 & British Journal of Mathematical and Statistical Psychology & \cite{wangMixturesT2024}\\
 & Communications for Statistical Applications and Methods & \cite{kyungBayesianAnalysisRandom2017}\\
 & Computational Statistics \& Data Analysis & \cite{martinez-vargasPottsCoxSurvivalRegression2023}\\
 & Electronic Journal of Statistics & \cite{barcellaBayesianNonparametricModel2016}\\
 & Journal of Computational and Graphical Statistics & \cite{mullerProductPartitionModel2011,muruaSemiparametricBayesianRegression2017,pageClusteringPredictionVariable2022,heinerProjectionApproachLocal2025}\\
 & Journal of Machine Learning Research & \cite{wadeImprovingPredictionDirichlet}\\
 & Journal of Statistical Software & \cite{grunFlexMixVersion22008,leischFlexMixGeneralFramework2004}\\
 & Journal of the American Statistical Association & \cite{daytonConcomitantVariableLatentClassModels}\\
 & Journal of the Royal Statistical Society Series C: Applied Statistics & \cite{garciaUnsupervisedBayesianClassification2024}\\
 & Scandinavian Journal of Statistics & \cite{quintanaClusterSpecificVariableSelection2015}\\
 & Statistical Methods in Medical Research & \cite{maIntegratingGenomicSignatures2018,proust-limaJointLatentClass2014,zhangJointLatentClass2022}\\
 & Statistical Modelling & \cite{jordanStatisticalModellingUsing2007}\\
 & Statistical Papers & \cite{songMultifeatureClusteringStep2024a}\\
 & Statistics and Computing & \cite{pageCalibratingCovariateInformed2018}\\
 & Statistics in Biosciences & \cite{leiJointAnalysisTwo2025,xuSubgroupBasedAdaptiveSUBA2016}\\
 & Statistics in Medicine & \cite{barcellaVariableSelectionCovariate2016,sunDirichletProcessMixture2017}\\
Biomedical/Public Health & BMC Medical Research Methodology & \cite{nguyenMultivariateLongitudinalData2023}\\
 & BMC Public Health & \cite{gartnerHowPredictiveFuture2024}\\
 & Clinical Pharmacology \& Therapeutics & \cite{dingEvaluatingPrognosticValue2024}\\
 & Environment International & \cite{liClusterbasedBaggingConstrained2019}\\
 & Journal of Cachexia, Sarcopenia and Muscle & \cite{choMetabolicPhenotypingComputed2024a}\\
 & Journal of Translational Medicine & \cite{palomino-echeverriaRobustClusteringStrategy2024}\\
 & Nature Communications & \cite{bayerNetworkbasedClusteringUnveils2025}\\
 & Oral Oncology & \cite{canahuateSpatiallyawareClusteringImproves2023a}\\
 & PLOS ONE & \cite{alexanderUsingTimeSeries2018}\\
 & Sleep & \cite{nasiriBoostingAutomatedSleep2021}\\
Computer Science & 2012 Conference on Technologies and Applications of Artificial Intelligence (TAAI) & \cite{gaoFeatureReductionText2012a}\\
 & ArXiv & \cite{trivediUtilityClusteringPrediction}\\
 & Artificial Intelligence in Medicine & \cite{esmailiMultichannelMixtureModels2021}\\
 & IEEE 26th International Conference on Intelligent Transportation Systems (ITSC) & \cite{almeidaLikelyLightAccurate2023}\\
 & IEEE Transactions on Mobile Computing & \cite{liExploitingComplexNetworkBased2024}\\
 & INFORMS Journal on Computing & \cite{chenClusterAwareSupervisedLearning2022}\\
 & International Conference on Emerging Research in Electronics, Computer Science and Technology & \cite{poojamrHybridDecisionSupport2015}\\
 & JMIR Medical Informatics & \cite{ramachandranAssessingValueUnsupervised2021}\\
 & Journal of Biomedical Informatics & \cite{eberhardDeepSurvivalAnalysis2024}\\
 & Knowledge and Information Systems & \cite{piernikStudyUsingData2021}\\
Engineering & Arabian Journal for Science and Engineering & \cite{mishraNovelVersionHorse2025}\\
 & Electronics & \cite{zhuShortTermLoadForecasting2024a}\\
 & Journal of Electrical Engineering & \cite{mokhtarAdaptiveBasedMachine2025a}\\
\end{longtable}

\begin{enumerate}[leftmargin=0pt, itemindent=1.5em, labelsep=0.5em, labelwidth=1em]

  \item Bayer et al. \cite{bayerNetworkbasedClusteringUnveils2025} used cluster membership indicators as covariates in a 
    Cox proportional hazards model for overall survival of 1,323 patients diagnosed with myeloid malignancies; that 
    is the clusters served as prognostic subgroups. In more detail, $k = 9$ patient clusters were derived using 
    a clustering method termed \emph{covariate-adjusted network clustering} (CANclust). At its core, CANclust 
    assumes a finite mixture model as in Equation~\eqref{eq:finite_mixture} for the covariates which captures dependencies among 
    certain mutational, cytogenetic and clinical covariates, while blocking confounding effects of demographic variables 
    like age and sex on clustering. The authors found that, compared to established risk 
   classifications, incorporating these clusters improved the predictive accuracy, with further gains observed when cluster assignments were combined with existing risk scores.
   In this example, clustering serves not only to identify subgroups but also to reduce dimensionality. Given the 
   high-dimensional space of potentially relevant covariates, some form of dimensionality reduction is typically required. 
   By summarising complex patterns among covariates into cluster assignments, this approach offers an alternative to 
   variable selection.

 \item Li et al. \cite{liClusterbasedBaggingConstrained2019} used clustering to predict nitrogen dioxide (NO$_2$) 
   and nitrogen oxide (NO$_\text{x}$) concentrations in the state of California. NO$_2$ and NO$_\text{x}$ measurements 
   were retrieved from air quality monitoring stations and field measurements. The data points were first clustered 
   based on geographical coordinates (latitude and longitude) using a combination of agglomerative clustering and $k$-means 
   which determines the number of clusters automatically and is referred to as the \emph{Kruskal-K-Mean method}. 
   Then within each cluster, additional clusters were obtained based on temporal characteristics of the measurements.
   The outcome model employed a so-called \emph{cluster-based bootstrap aggregating} (bagging) strategy, 
   where units were systematically excluded and stratified bootstrap sampling 
   was performed on the remaining clusters. A base model consisting of a 
   mixed-effects model was fitted to each bootstrap sample. This process was repeated for each cluster resulting in 
   predictions for the excluded samples. The individual predictions from each bootstrap sample were combined into one ensemble prediction.

\item In a study of metabolic syndrome, osteoporosis, and sarcopenia, Cho et al. \cite{choMetabolicPhenotypingComputed2024a} first used 
  hierarchical clustering of body composition variables derived from computer tomography to identify three metabolic 
  phenotypes (normal, metabolic syndrome, and osteosarcopenia), followed by comparisons of clinical characteristics 
  including age, visceral fat area and density, subcutaneous fat area and density, skeletal muscle area and density,
  across clusters (i.e., the outcome modelling part).

\item Palomino-Echeverria et al.\cite{palomino-echeverriaRobustClusteringStrategy2024} developed a clustering framework called ClustALL that aims to identify patient 
    clusters (referred to as stratifications) robust to both variations in the study population and differences in 
    the parameter settings of the clustering procedure. The framework comprises three steps. In Step~1, 
    hierarchical clustering is applied to covariates rather than to units (i.e., patients), producing a dendrogram in which covariates 
    are grouped based on their similarity. At each height of the dendrogram, the variables belonging 
    to the clusters formed at that level are transformed using PCA, with the first 
    three principal components retained (singletons are kept as-is). 
    This yields a reduced-dimension representation of the data, termed an \emph{embedding}, for each dendrogram height. 
    In Step~2, two dissimilarity matrices between patient pairs are computed per embedding - one based on correlation 
    distance, the other based on Gower's \cite{gowerGeneralCoefficientSimilarity1971} distance. Hierarchical clustering and $k$-means are applied to the former, 
    and hierarchical clustering and $k$-medoids to the latter. In Step~3, Jaccard similarity indices based on 
    1,000 bootstrap samples are calculated for each clustering from Step~2, and those with an index below 85\% 
    are excluded. The remaining set of partitions of $[n]$ are termed \emph{population-based robust stratifications}. Jaccard 
    distances (i.e., dissimilarity measures) among these are then computed, and from each group of similar stratifications a representative 
    is selected, yielding \emph{parameter-based robust stratifications}. Applied to 766 patients with acutely 
    decompensated liver cirrhosis, ClustALL identified five such representative stratifications - i.e., five different 
    representative partitions of the index set $\{1,\ldots,766\}$. One of these stratifications divided the patients into 
    three clusters, and these clusters were subsequently used as prognostic subgroups in cumulative incidence 
    analyses for acute-on-chronic liver failure (ACLF) and death, which comprises the outcome modelling step.

  \item Ding et al.\ \cite{dingEvaluatingPrognosticValue2024} applied
    $k$-means clustering to functional principal component scores obtained from longitudinal lactate dehydrogenase (LDH) 
    measurements in patients with metastatic colorectal cancer (mCRC). Using data from two phase III clinical 
    trials ($n=824$ and $n=210$), three patient clusters were retained, which then served as prognostic subgroups in 
    survival analysis. 

  \item Nasiri and Clifford \cite{nasiriBoostingAutomatedSleep2021} aimed to classify sleep stages using 
    EEG recordings obtained from six channels. To this end, 794 patients were clustered by applying spectral clustering 
    to the geometric means of the covariance matrices between the EEG sites obtained over time (each covariance matrix 
    is based on a 30 second time window, with $30 \times f_s$ EEG measurements taken, where $f_s$ denotes the sampling frequency).
    This resulted in the partition of the training data into five clusters. The covariance matrices for each 30-second epoch 
    were then converted to a feature vector. A convolutional neural network was fitted 
    separately for each cluster using these features as inputs 
    and five different sleep stages (Wake, N1, N2, N3, REM) as the outcomes to be predicted. 
    This clustering-based approach outperformed a single convolutional neural network trained on the entire training data across multiple 
    prediction metrics, as the 
    five convolutional neural networks were each trained on more homogeneous data.

  \item Canahuate et al. \citep{canahuateSpatiallyawareClusteringImproves2023a} proposed a two-stage clustering 
  framework applied to 575 head and neck cancer patients. First, $k$-means clustering partitioned patients 
  into three groups based on radiomic (RM) features, while hierarchical clustering 
  produced two clusters from lymph node (LN) involvement variables. The RM and 
  LN cluster memberships were then combined to define three risk groups, which were then incorporated 
  as covariates (i.e., as prognostic groups) in Cox proportional hazards models for survival outcomes.

\item Gartner et al.\ \cite{gartnerHowPredictiveFuture2024} identified 10 clusters among 412,997 patients using 8 healthcare 
  utilisation and comorbidity variables. Adding cluster membership indicators to logistic regression models improved the area under the curve (AUC) for 
  predicting emergency admissions, accident and emergency attendance, general practitioner contacts, and mortality compared to models without them.

\item Nguyen et al.\ \cite{nguyenMultivariateLongitudinalData2023} used data from the CARDIA cohort, comprising 3,539 individuals enrolled between 
  1985 and 1986. The aim was to assess and compare methods that include longitudinal covariates to predict the time 
  until the occurrence of a cardiovascular event. 
  Thirty-five variables were measured at six visits over 15 years (at years 0, 2, 5, 7, 10, and 15), and 
  participants were followed for cardiovascular events during the subsequent 17 years. The authors compared six strategies 
  for incorporating longitudinal measurements into a survival model, including one approach that first clustered patients 
  based on the trajectory of each variable and then used the resulting cluster membership indicators as covariates in survival models
  such as random survival forests and Cox proportional hazard models. The longitudinal covariates were clustered 
  using a three-step procedure first described by Leffondre \cite{leffondreStatisticalMeasuresWere2004} implemented 
  in \emph{R} via the \emph{traj}-package \cite{sylvestre2025} . In step 1, 24 characteristics (called "measures of change") 
  of the longitudinal covariate were extracted for each participant in the study. 
  Step 2 applied principal components analysis 
  to these characteristics and step 3, performed $k$-means or $k$-medoids clustering on the resulting principal components 
  scores. This approach effectively treats clusters as subgroups.

\item Alexander et al. \cite{alexanderUsingTimeSeries2018, alexanderIntegratingDataRandomized2017} 
used clustering as a preprocessing step before matching patients from the 
treatment arms of randomized controlled trials (RCTs, $n = 1320$) with 
patients from an observational study (OS, $n = 2642$) to predict 
trajectories of patient-reported pain scores in patients with painful 
diabetic peripheral neuropathy. The OS patients were first divided into 
six clusters using hierarchical clustering on nine baseline covariates. 
Then, within each OS cluster, coarsened exact matching was applied 
to identify RCT patients sharing similar coarsened covariate profiles. 
Separate penalized time-series regression models with lagged covariates 
were then fitted for each cluster. Predictions for a new patient were 
obtained by first assigning the patient to a cluster using an ensemble 
of $k$-Nearest Neighbours and supervised Fuzzy C-Means, then using the 
cluster-specific model to simulate virtual patients with possible pain score 
trajectories.
    
\end{enumerate}

\newpage 

\section{Discussion} \label{sec:discussion}
Our primary objective was to provide an overview of existing 
approaches that use clustering of observational 
units based on covariates to model an outcome of interest, with particular emphasis on 
medical and clinical applications. We identified 55 records primarily 
through Web of Science and PubMed, spanning statistics, biomedical 
sciences and public health, computer science, and engineering, with 
biomedical and public health applications clearly predominating.

We distinguished between informed-cluster and agnostic-cluster
models. Informed-cluster models (PPM/PPMx, FMR, CluSL) 
incorporate outcome information into the clustering process itself, 
while agnostic-cluster models employ a two-step procedure in which 
observational units are first clustered based solely on covariates, 
then clustering-derived variables are used in the outcome model. 
These approaches were clearly separated by disciplinary context, 
with informed-cluster models appearing almost exclusively in 
methodological works published in statistical journals, while 
records published in biomedical and public health journals exclusively employed 
agnostic-cluster models.

The most common objective for employing a clustering-based outcome model was subgroup identification, followed by 
dimensionality reduction and the generation of new features or covariates (referred to as feature extraction in Table~\ref{tbl:all_data}). 
These goals can all be viewed in light of another goal listed in Table~\ref{tbl:all_data}, namely that of addressing heterogeneity.

If the outcome model is not sufficiently flexible, for example if it is restricted to linear main effects and omits important 
interactions or nonlinearities, predictions may be systematically biased. Increasing flexibility can reduce this bias but 
can also increase the out-of-sample prediction error due to overfitting. Thus, to minimize the prediction error this trade-off 
has to be considered \citep[][]{hastie2009elements} (Chapter 7.3). Outcome models using clustering are one approach to address 
this. As the clustering step is highly flexible in the covariate space, the subsequent outcome model remains parsimonious 
by using a low-dimensional set of cluster-derived summaries (e.g., cluster membership indicators, cluster assignment probabilities, 
dissimilarities to centroids) rather than all covariates individually. This can reduce overfitting when high-dimensional covariate information would otherwise 
increase the out-of-sample prediction error, especially when clustering does not use information on the outcome. However, 
this strategy can only improve predictions if the clusters defined in the covariate space correspond to meaningful differences 
in the outcome distribution. Otherwise, cluster-derived variables will only add noise. Informed-cluster models incorporate the outcome into clustering and therefore can improve this alignment but in this review we did not identify studies that compare the performance with agnostic-cluster models.

Note that clustering-based outcome models that fail to recover a true underlying subgroup structure may still achieve  
adequate predictive performance. For instance, if there are no true subgroups in the covariate space that are related to the outcome, 
adding clustering information increases model complexity but will affect prediction only through increased variance and a higher risk of 
overfitting.  In contrast, when the objective is to identify subgroups of patients with  differential outcomes or treatment 
effects (the most prevalent objective in our review), the derived clusters need to correspond to the underlying subgroups.

Subgroup identification adds further inferential aspects to clustering-based outcome models. In addition to  
outcome modelling, the existence, definition and number of subgroups in the population can be investigated. 
In all models reviewed here, except for PPM/PPMx, a single clustering solution is used as a plug-in estimate of this subgroup 
structure and the uncertainty of these estimates is not quantified. This uncertainty depends on the sample size, the 
number of covariates and on how well separated the true subgroups are with respect to the variables used for clustering \citep{dalmaijerStatisticalPowerCluster2022}.

A particularly interesting application of clustering as a 
dimensionality reduction tool is the incorporation of longitudinal or functional covariates into prediction models. Here clustering groups observations based on individual trajectories of covariate values and naturally accommodates differences in the number of 
measurements between units.

Clustering-based outcome models can be applied in rare disease research, where heterogeneity is  a central characteristic 
\cite{morelMeasuringWhatMatters2017, 
murrayApproachesAssessmentClinical2023}, particularly when high dimensional baseline data is available.
While rare disease studies need not be small \citep{hee2017does}, in many settings large prospective studies are not feasible. 
However, if larger cross-sectional historic data sets containing covariate information (but not the outcome) are available 
through registries or electronic health records, cluster definitions can be derived from  these historical data sets and then used 
to fit an outcome model in a prospective longitudinal study with outcome data. 
Ma et al. \citep{maIntegratingGenomicSignatures2018} 
demonstrated this approach by using clustering to derive power priors, while Alexander et al. 
\citep{alexanderUsingTimeSeries2018} combined samples from observational studies and RCTs via coarsened exact matching 
within clusters.

There is also potential for such methods in the analysis of controlled clinical trials evaluating new therapies. In particular,
clustering based outcome models can be used for covariate adjustment to improve power of statistical tests and to increase the 
precision of treatment effect estimates. The outcome model can be either derived from historical data and then used in the 
clinical trial as a super-covariate \cite{Holzhauer2023}, or it can be fitted using the clinical trial data.  In the latter case, 
treatment is an additional covariate in the model and must not be used for clustering.  In addition, as discussed above, a hybrid 
approach performs only the clustering step using the historical data and estimates the outcome model with the trial data.
Another application is to assess treatment effect heterogeneity by including treatment-cluster interactions to estimate subgroup 
specific treatment effects, with subgroups defined by the identified clusters. 

This review has several limitations. 
As is typical for scoping reviews, we aimed to describe and classify methods rather than to provide a quantitative comparison 
of predictive performance across studies.
The methods were identified in the  Web of Science and PubMed databases, supplemented by a small number of manually added records. 
The identification of methods depend on specific search strings and inclusion and exclusion criteria. These were chosen to 
include relevant manuscripts while keeping the screening workload manageable. For example, we excluded work focused on imaging data. 
As “clustering-based prediction model” is not a standard term, and related approaches are described under a range of labels in the 
literature (see, for example, the terminology used for informed-cluster models in Section~\ref{sec:key_approaches}) we may have missed 
relevant methods that fall within the intended scope despite broad search strings.

To summarise, this scoping review shows that clustering approaches are mainly used to define covariate-based 
patient subgroups or to aggregate covariate information to be used in outcome models. In clinical applications, the 
primary aim is interpretable risk stratification. For the use of these methods in prospective clinical trials explainable 
clusters with demonstrated cluster stability will be important. 

\section*{Funding}
This work was performed within the project INVENTS, which has received funding 
from the Horizon Europe Research and Innovation programme under grant agreement 
101136365, the Swiss State Secretariat for Education, Research and Innovation 
(SERI) and by UKRI Innovation UK under their Horizon Europe Guarantee scheme.
Views and opinions expressed are those of the author(s) only and do not necessarily reflect those of the European Union or the
European Health and Digital Executive Agency. Neither the European
Union nor the granting authority can be held responsible for them.

\section*{Consent for Publication}
All authors consent to the publication of this manuscript.

\section*{Data Availability}
The datasets generated during and/or analyzed during the current study are available as part
of the supplementary materials of this manuscript.

\newpage
\clearpage
\bibliography{references.bib}

\newpage
\appendix
\renewcommand{\thetable}{A\arabic{table}}
\setcounter{table}{0}
\clearpage
\phantomsection

\section{Appendix}
\label{app:tables}

\begin{longtable}[t]{>{\raggedright\arraybackslash}p{4cm}cc>{\raggedright\arraybackslash}p{6cm}}
\caption{List of all records included in the scoping review. Data Base denotes the data base through which the record was identified (WoS = Web of Science)}
\label{tbl:studies}\\
\toprule
Record & Year & Data Base & Source\\
\midrule
\endfirsthead
\multicolumn{4}{@{}l}{\textit{(continued)}}\\
\toprule
Record & Year & Data Base & Source\\
\midrule
\endhead

\endfoot
\bottomrule
\endlastfoot
Dayton and Macready \cite{daytonConcomitantVariableLatentClassModels} & 1988 & Other & Journal of the American Statistical Association\\
Leisch \cite{leischFlexMixGeneralFramework2004} & 2004 & Other & Journal of Statistical Software\\
Jordan, Livingstone, and Barry \cite{jordanStatisticalModellingUsing2007} & 2007 & WoS & Statistical Modelling\\
Grün and Leisch \cite{grunFlexMixVersion22008} & 2008 & Other & Journal of Statistical Software\\
Proust-Lima and Taylor \cite{proust-limaDevelopmentValidationDynamic2009} & 2009 & WoS & Biostatistics\\
Müller, Quintana, and Rosner \cite{mullerProductPartitionModel2011} & 2011 & WoS & Journal of Computational and Graphical Statistics\\
Gao and Chien \cite{gaoFeatureReductionText2012a} & 2012 & WoS & 2012 Conference on Technologies and Applications of Artificial Intelligence (TAAI)\\
Müller et al. \cite{mullerBayesianInferenceLongitudinal2014} & 2014 & WoS & Biostatistics\\
Proust-Lima et al. \cite{proust-limaJointLatentClass2014} & 2014 & Other & Statistical Methods in Medical Research\\
Wade et al. \cite{wadeImprovingPredictionDirichlet} & 2014 & WoS & Journal of Machine Learning Research\\
Page and Quintana \cite{pagePredictionsBasedClustering2015} & 2015 & WoS & Bayesian Analysis\\
Pooja and Pushpalatha \cite{poojamrHybridDecisionSupport2015} & 2015 & WoS & International Conference on Emerging Research in Electronics, Computer Science and Technology\\
Quintana, Müller, and Papoila \cite{quintanaClusterSpecificVariableSelection2015} & 2015 & WoS & Scandinavian Journal of Statistics\\
Trivedi, Pardos, and Heffernan \cite{trivediUtilityClusteringPrediction} & 2015 & Other & ArXiv\\
Barcella et al. \cite{barcellaBayesianNonparametricModel2016} & 2016 & WoS & Electronic Journal of Statistics\\
Barcella et al. \cite{barcellaVariableSelectionCovariate2016} & 2016 & WoS & Statistics in Medicine\\
Xu et al. \cite{xuSubgroupBasedAdaptiveSUBA2016} & 2016 & WoS & Statistics in Biosciences\\
Kyung \cite{kyungBayesianAnalysisRandom2017} & 2017 & WoS & Communications for Statistical Applications and Methods\\
Murua and Quintana \cite{muruaSemiparametricBayesianRegression2017} & 2017 & WoS & Journal of Computational and Graphical Statistics\\
Sun et al. \cite{sunDirichletProcessMixture2017} & 2017 & WoS & Statistics in Medicine\\
Keller et al. \cite{kellerCovariateadaptiveClusteringExposures2017} & 2017 & PubMed & Annals of Applied Statistics\\
Ma, Hobbs, and Stingo \cite{maIntegratingGenomicSignatures2018} & 2018 & PubMed & Statistical Methods in Medical Research\\
Page and Quintana \cite{pageCalibratingCovariateInformed2018} & 2018 & WoS & Statistics and Computing\\
Gronsbell et al. \cite{gronsbellAutomatedFeatureSelection2019} & 2018 & PubMed & Biometrics\\
Alexander et al. \cite{alexanderUsingTimeSeries2018} & 2018 & PubMed & PLOS ONE\\
Li et al. \cite{liClusterbasedBaggingConstrained2019} & 2019 & PubMed & Environment International\\
Lijoi, Prünster, and Rigon \cite{lijoiPitmanYorMultinomial2020} & 2020 & WoS & Biometrika\\
Esmaili et al. \cite{esmailiMultichannelMixtureModels2021} & 2021 & PubMed & Artificial Intelligence in Medicine\\
Nasiri and Clifford \cite{nasiriBoostingAutomatedSleep2021} & 2021 & PubMed & Sleep\\
Page, Quintana, and Rosner \cite{pageDiscoveringInteractionsUsing2021} & 2021 & WoS & Annals of Applied Statistics\\
Piernik and Morzy \cite{piernikStudyUsingData2021} & 2021 & WoS & Knowledge and Information Systems\\
Ramachandran et al. \cite{ramachandranAssessingValueUnsupervised2021} & 2021 & PubMed & JMIR Medical Informatics\\
Chen and Xie \cite{chenClusterAwareSupervisedLearning2022} & 2022 & WoS & INFORMS Journal on Computing\\
Page, Quintana, and Müller \cite{pageClusteringPredictionVariable2022} & 2022 & WoS & Journal of Computational and Graphical Statistics\\
Zhang and Simonoff \cite{zhangJointLatentClass2022} & 2022 & PubMed & Statistical Methods in Medical Research\\
de Almeida and Mozos \cite{almeidaLikelyLightAccurate2023} & 2023 & WoS & IEEE 26th International Conference on Intelligent Transportation Systems (ITSC)\\
Canahuate et al. \cite{canahuateSpatiallyawareClusteringImproves2023a} & 2023 & PubMed & Oral Oncology\\
Martinez-Vargas and Murua-Sazo \cite{martinez-vargasPottsCoxSurvivalRegression2023} & 2023 & WoS & Computational Statistics \& Data Analysis\\
Nguyen et al. \cite{nguyenMultivariateLongitudinalData2023} & 2023 & PubMed & BMC Medical Research Methodology\\
Heiner, Page, and Quintana \cite{heinerProjectionApproachLocal2025} & 2023 & WoS & Journal of Computational and Graphical Statistics\\
Wang et al. \cite{wangMixturesT2024} & 2023 & PubMed & British Journal of Mathematical and Statistical Psychology\\
Argiento et al. \cite{argientoClusteringBloodDonors2024} & 2024 & WoS & Biometrics\\
Cho et al. \cite{choMetabolicPhenotypingComputed2024a} & 2024 & PubMed & Journal of Cachexia, Sarcopenia and Muscle\\
Ding et al. \cite{dingEvaluatingPrognosticValue2024} & 2024 & PubMed & Clinical Pharmacology \& Therapeutics\\
Eberhard et al. \cite{eberhardDeepSurvivalAnalysis2024} & 2024 & PubMed & Journal of Biomedical Informatics\\
Garcia et al. \cite{garciaUnsupervisedBayesianClassification2024} & 2024 & PubMed & Journal of the Royal Statistical Society Series C: Applied Statistics\\
Gartner et al. \cite{gartnerHowPredictiveFuture2024} & 2024 & PubMed & BMC Public Health\\
Li et al. \cite{liExploitingComplexNetworkBased2024} & 2024 & WoS & IEEE Transactions on Mobile Computing\\
Palomino-Echeverria et al. \cite{palomino-echeverriaRobustClusteringStrategy2024} & 2024 & WoS & Journal of Translational Medicine\\
Zhu et al. \cite{zhuShortTermLoadForecasting2024a} & 2024 & WoS & Electronics\\
Song et al. \cite{songMultifeatureClusteringStep2024a} & 2024 & WoS & Statistical Papers\\
Lei et al. \cite{leiJointAnalysisTwo2025} & 2025 & WoS & Statistics in Biosciences\\
Mishra, Mahajan, and Garg \cite{mishraNovelVersionHorse2025} & 2025 & WoS & Arabian Journal for Science and Engineering\\
Mokhtar \cite{mokhtarAdaptiveBasedMachine2025a} & 2025 & WoS & Journal of Electrical Engineering\\
Bayer et al. \cite{bayerNetworkbasedClusteringUnveils2025} & 2025 & PubMed & Nature Communications\\
\end{longtable}
\newpage 

\begin{table}
  \centering
  \caption{Search strings for Web of Science and PubMed and prompts used.}
  \label{tbl:search}
  \begin{tabular}{p{3cm} p{6cm} p{3cm} p{2cm}}
    \toprule
    Database        & Search string & Number of Results & Date Last Searched\\
    \midrule
    Web of Science  & ALL=("covariate clustering" OR "feature clustering" OR "random partition" OR
"clustering of covariates" OR "clustering for feature extraction" OR
"covar* NEAR partition" OR "feature* NEAR partition")
    AND ALL=("prediction" OR "classification" OR "regression")& 481 & 2025-07-11\\
    PubMed & (covariate clustering[Title/Abstract] OR feature clustering[Title/Abstract]) AND (prediction[Title/Abstract] OR classification[Title/Abstract]) & 252 &  2025-07-11\\
    \bottomrule
        \end{tabular}
    \end{table}
\newpage

\end{document}

\maketitle

\newpage 

\section{List of abbreviations}
\begin{table}[htbp]
\centering
\begin{tabular}{ll}
\hline
\textbf{Abbreviation} & \textbf{Definition} \\
\hline
ACLF & Acute-on-chronic liver failure \\
AI & Artificial Intelligence \\
AIC & Akaike's Information Criterion \\
AUC & Area under the curve \\
BIC & Bayesian Information Criterion \\
CEM & Coarsened exact matching \\
EEG & Electroencephalography \\
FMR & Finite mixtures of regression models \\
HAM-D & Hamilton Depression Rating Scale \\
HTE & Heterogeneity of treatment effects \\
LASSO & Least absolute shrinkage and selection operator \\
LCA & Latent class analysis \\
LDH & Lactate dehydrogenase \\
LLM & Large language model \\
LN & Lymph node \\
LUSC & Lung squamous cell carcinoma \\
NO2 & Nitrogen dioxide \\
PPM & Product partition models \\
PPMx & Product partition models with covariates \\
PRISMA & Preferred Reporting Items for Systematic Reviews and Meta-Analyses \\
PSA & Prostate-Specific Antigen \\
RCT & Randomized Controlled Trial \\
REM & Rapid eye movement (sleep) \\
RM & Radiomic features \\
UCI & University of California, Irvine \\
WBC & White blood cell counts \\
ZIP & Zero-inflated Poisson \\
\hline
\end{tabular}
\end{table}

\newpage 

\section{Preregistration protocol and extraction form}

This section is available as a separate document at \url{https://osf.io/e6k8f/files/rsfh4}.
In the following, details on the literature review are provided. We follow the structure of the PROSPERO systematic review registration. Sections of the PROSPERO guidelines that are not applicable are omitted.

\subsection{Citation} 
\textit{This section has been removed to ensure anonymisation during peer review. Full details, including pre-registration documents, will be made available upon acceptance.}

\subsection{Review question}

The objective of this review is to identify statistical approaches that use clustering of observational units (e.g., patients) based on baseline covariates and then use these clusters in the development of prediction models for an endpoint/outcome of interest. Of interest are methods used in clinical trials for outcome prognosis and to predict treatment effects based on covariates, and methods where the baseline covariates are longitudinal measurements. Thus, the term "baseline" will include variables covering the longitudinal history of covariates, e.g., the BMI measured every two months over the last two years. 

\subsection{Searches} 

A comprehensive literature search will be conducted in the databases
\emph{Web of Science} (WoS) and \emph{PubMed}. The search terms included are listed in 
Table~\ref{tbl-search_terms}. In addition, articles identified by manual search will be included if they 
meet the inclusion and exclusion criteria.

\begin{table}[htbp]
\centering
\caption{Search terms for the literature review.}
\label{tbl-search_terms}
\begin{tabular}{p{3cm}p{12cm}}
\toprule
\textbf{Database} & \textbf{Search string} \\
\midrule
Web of Science & ALL=(``covariate clustering'' OR ``feature clustering'' OR ``random partition'' OR ``clustering of covariates'' OR ``clustering for feature extraction'' OR ``covar* NEAR partition'' OR ``feature* NEAR partition'') AND ALL=(``prediction'' OR ``classification'' OR ``regression'') \\
PubMed & (covariate clustering[Title/Abstract] OR feature clustering[Title/Abstract]) AND (prediction[Title/Abstract] OR classification[Title/Abstract]) \\
\bottomrule
\end{tabular}
\end{table}

\subsection{ Types of studies to be included }

The review will be restricted to methodological papers that describe the proposed methods in sufficient detail to allow implementation and assessment of their performance through simulations, applications to benchmark datasets, or analytical evaluation.

Only articles that satisfy all inclusion and none of the exclusion criteria will be included in the review.

\subsubsection{Inclusion criteria}

\begin{itemize}
  \item A method to fit prediction models employing clustering of observational units (e.g., patients, subjects) based on covariates is proposed or investigated.
  \item The proposed methods are described in sufficient detail to allow for an independent implementation
\end{itemize}

\subsubsection{Exclusion criteria}

\begin{itemize}
 \item Applied publications where clustering algorithms are applied but no methodology is proposed or described in sufficient detail.
 \item Clustering algorithms proposed without application as covariates in a prediction model.
 \item The method is based on clustering the covariates themselves (e.g., to reduce the dimensionality of the covariate space) as opposed to observational units. 
 \item Publications that use clustering to analyze or process imaging data
\end{itemize}

\subsection{Condition or domain being studied}

Statistical methods, clustering of covariates in prediction models, with
a particular emphasis on the analysis of clinical trial data.

\subsection{Main outcome(s)}

This review focuses on statistical methods for clustering of observational units based on covariates
within prediction models, with the primary objective of summarizing the
current range of approaches developed for this purpose together with their statistical properties and known underlying assumptions and limitations.

\subsection{Data Extraction (selection and coding)}

The initial abstract screening of the search results for relevant articles will be done by two reviewers independently. If the number of search results is too high, a risk-based approach will be implemented with one reviewer per abstract and only a random subset of abstracts will be reviewed by a second reviewer. If the number of discrepancies is considered too high, additional training and modification of the extraction form will be provided. Titles and abstracts will be screened and
papers that clearly do not meet the inclusion criteria will be excluded.
Any disagreements will be resolved in discussion or re-screened by a
third reviewer. For all relevant abstracts, full texts will be obtained
and reviewed according to the prespecified inclusion criteria. From the
full text, data on the proposed methods and, if applicable, on
simulation studies or worked analysis examples will be collected by two
independent reviewers. Characteristics on the proposed methods will
include information on whether the methods used purely algorithmic
clustering methods, model-based clustering methods, or hybrid clustering
methods. The list of characteristics to be extracted from the full text are given in \autoref{tab:extract}. \autoref{tab:extract} will also be used as an extraction form. We will extract characteristics pertaining to bibliographic information, information on the proposed methods and details on 
how these methods are evaluated in simulation studies or real-world data settings. Depending on the number of abstracts identified for full data extraction, the review will be supported by large language models (LLM) for reviewing such as Elicit. If LLMs will be used for certain variables, a proportion of at least $5\% $ of the papers will be reviewed manually. If the error rate is too high, the critical results results derived with LLM will be manually checked.
Based on a pilot of 10 full papers review, the data extraction form will be modified to address the learnings. The changes and potential use of LLM will be specifically noted in the review report. It will be reported which data have been derived automatically (such as bibliographic information from registers or using LLM) and which have been derived manually.

\newpage

\begin{longtable}{p{8cm}p{7.5cm}}
\caption{Data extraction form.}\label{tab:extract} \\
\toprule
\textbf{Information to Extract} & \textbf{Further Details} \\
\midrule
\endfirsthead
\toprule
\textbf{Information to Extract} & \textbf{Further Details} \\
\midrule
\endhead
\midrule
\multicolumn{2}{r}{\textit{Continued on next page}} \\
\endfoot
\bottomrule
\endlastfoot
Bibliographic Information & Authors, Publication Year, DOI, Title, Journal, Volume Pages \\
Proposed Method & Brief description of the method (1 or 2 sentences) \\
Which clustering methods are employed & e.g., $k$-means, hierarchical clustering \\
Are covariates assumed to be cross-sectional or longitudinal or not specified any further? & \\
What is the goal of clustering? & e.g., dimensionality reduction, covariate augmentation \\
Algorithmic vs model-based clustering & Pertains only to the clustering method, not the model for the endpoint. Algorithmic clustering methods do not (explicitly) assume a probability model for the covariates. Examples of algorithmic clustering methods are $k$-means or hierarchical clustering. Model-based clustering methods assume a probability model for the covariates. Most commonly finite-mixture models are assumed. \\
Classification or prediction task & Is the endpoint categorical (i.e., classification task) or metric (i.e., prediction task) \\
Does the endpoint of interest inform the clustering? & Some methods derive the clusters by including the endpoint (supervised learning, e.g., \cite{andradeConvexCovariateClustering2021}), others derive the clusters based on the covariates only (unsupervised learning, e.g., \cite{piernikStudyUsingData2021}) \\
Are the methods used to predict the endpoint or the treatment effect (or both) & \\
How is the number of clusters determined? & Many clustering algorithms require that the number of clusters be set a-priori. How does the proposed method handle this issue? \\
Simulation study (yes/no) & \\
Assumed sample sizes (range) in the simulation study & Minimum and Maximum \\
Number of covariates considered (range) in the simulation study & Minimum and Maximum \\
Performance measures used in simulation study (if simulation = yes) & \\
Real-world or benchmark data example (yes/no) & e.g., the University of California Irvine Machine Learning data sets are used in several publications \\
Sample sizes (range) in the real-world data & Minimum and Maximum \\
Number of covariates in the real-world data & Minimum and Maximum \\
Performance measures used in real-world example & e.g., performance on test set \\
Domain(s) of real-world data example(s) & e.g., Medicine, Biology, Text Classification, etc. \\
Software available (y/n)? & \\
Is software open source (y/n)? & \\
Software implemented in R, Python or another? & \\
Link to software & provide URL \\
Included in review (y/n) & \\
Reason for exclusion (if study included in review = no) & applied publication, clustering method only, other (see exclusion criteria above) \\
\end{longtable}

\subsection{Review team members and their affilliations (alphabetic order)}
\textit{This section has been removed to ensure anonymisation during peer review.}
Additional reviewers might be added depending on the number of abstracts/ full papers being identified.

\subsection{Type and method of review}\label{type-and-method-of-review}

Methodology, Systematic Review

\subsection{Anticipated or actual start date}

01 April 2025

\subsection{Anticipated Completion}

30 September 2025

\subsection{Funding sources/sponsors }

\subsection{Conflicts of interest}\label{conflicts-of-interest}

None.

\subsection{Language}

English

\newpage 

\section{Elicit prompts}

\begin{longtable}{p{4cm}p{10cm}}
\hline
\textbf{Characteristic} & \textbf{Prompt Used} \\
\hline
\endfirsthead

\hline
\textbf{Characteristic} & \textbf{Prompt Used} \\
\hline
\endhead

\hline
\endfoot

\hline
\endlastfoot

Bibliographic Information & Extract or find the following bibliographic information:
\begin{itemize}[nosep,leftmargin=*]
  \item Authors
  \item Title
  \item DOI
\end{itemize} \\
Decision to include & If the publication proposes or investigates a method to fit a prediction model employing clustering of observational units (e.g., patients, subjects) based on covariates and if the method is sufficiently described such that it can be implemented independently return ``Include''. Otherwise return ``Exclude''. If the publication is about an application without introducing a new methodology or if just a clustering algorithm without using clustering of covariates in a prediction model or if the method clusters covariates themselves and not observational units or if the publication is about imaging data return ``Exclude''. If ``Exclude'' provide a reason. \\
Description of method & Briefly describe how clustering based on covariates is used to build a prediction model. \\
Clustering method used & Name the method used to cluster the observations based on covariates (e.g., k-means, hierarchical clustering, finite mixture models). \\
Time-varying covariates & Does the method assume that the covariates used for clustering are cross-sectional only, longitudinal only, or is it general enough to allow for both cross-sectional and longitudinal covariates? \\
Goal of clustering & What is the purpose of clustering? Select all that apply:
\begin{itemize}[nosep,leftmargin=*]
  \item Dimensionality reduction
  \item Feature extraction/creation
  \item Handling missing data
  \item Subgroup identification
  \item Other (specify)
\end{itemize} \\
Model-based vs. algorithmic clustering & Is the clustering method algorithmic or model-based?
\begin{itemize}[nosep,leftmargin=*]
  \item Algorithmic (no probability model needed)
  \item Model-based (assumes probability distribution)
  \item Hybrid approach
  \item Other/Unclear (explain)
\end{itemize}
Note: Algorithmic methods use distance/similarity measures without assuming a specific probability distribution. Model-based methods assume the data follows a particular probability distribution and estimate parameters of that distribution. \\
Outcome scale & What type of endpoints does the method handle?
\begin{itemize}[nosep,leftmargin=*]
  \item Binary/categorical only
  \item Continuous only
  \item Both categorical and continuous
  \item Other (specify)
\end{itemize} \\
Outcome-informed vs. outcome-agnostic & Does the outcome variable inform the clustering?
\begin{itemize}[nosep,leftmargin=*]
  \item No (unsupervised clustering)
  \item Yes (supervised clustering)
  \item Unclear from the paper
\end{itemize}
Note: Unsupervised clustering uses only input variables/covariates to form clusters, ignoring the outcome variable during cluster formation. Supervised clustering incorporates the outcome variable during the clustering process. \\
Treatment effect involved & What does the method predict?
\begin{itemize}[nosep,leftmargin=*]
  \item Outcome/endpoint only
  \item Treatment effect only
  \item Both outcome and treatment effect
  \item Other (specify)
\end{itemize} \\
Number of clusters & How is the number of clusters determined?
\begin{itemize}[nosep,leftmargin=*]
  \item BIC
  \item Fixed in advance
  \item Other (specify)
\end{itemize} \\
Simulation studies & Does the study include a simulation component?
\begin{itemize}[nosep,leftmargin=*]
  \item Yes
  \item No
\end{itemize}
If yes, provide simulation details:
\begin{itemize}[nosep,leftmargin=*]
  \item Minimum sample size
  \item Maximum sample size
  \item Minimum number of covariates
  \item Maximum number of covariates
  \item Performance measures used to evaluate the method
\end{itemize} \\
Real-world data examples & Does the publication provide an application to real-world data?
\begin{itemize}[nosep,leftmargin=*]
  \item Yes
  \item No
\end{itemize}
If yes, provide real-world data details:
\begin{itemize}[nosep,leftmargin=*]
  \item Minimum sample size
  \item Maximum sample size
  \item Minimum number of covariates
  \item Maximum number of covariates
  \item Performance measures used to evaluate the method using real-world data
  \item Domain of the real-world data (e.g., medicine, biology)
\end{itemize} \\
Software & Is software available?
\begin{itemize}[nosep,leftmargin=*]
  \item Yes
  \item No
  \item Unclear/Not mentioned
\end{itemize}
If yes: Is the software open source?
\begin{itemize}[nosep,leftmargin=*]
  \item Yes
  \item No
  \item Unclear
\end{itemize}
Software implemented in (select all that apply):
\begin{itemize}[nosep,leftmargin=*]
  \item R
  \item Python
  \item C/C++
  \item Other (specify)
\end{itemize}
Link to software: \\
\end{longtable}

\newpage 

\section{Characteristics by record}

See accompanying CSV file. 

\blandscape
\section{Goals and methods for selecting the number of clusters by discipline}

\begin{table}
  \centering
\begin{tabular}{lcccc}
\toprule
  & Biomedical/Public Health & Computer Science & Engineering & Statistics\\
\midrule
Objective &  &  &  & \\
Subgroup Identification & 5 & 5 & 2 & 19\\
Dimensionality Reduction & 8 & 4 & 0 & 9\\
Feature Extraction & 6 & 7 & 1 & 6\\
Addressing Heterogeneity & 0 & 1 & 0 & 8\\
Improve Prediction & 1 & 2 & 0 & 3\\
Include Longitudinal Covariates & 0 & 0 & 0 & 3\\
Account for Missing Data & 0 & 0 & 0 & 1\\
Variable Selection & 0 & 0 & 0 & 1\\
Cluster Number Method & & & & \\
AIC & 1 & 2 & 0 & 2\\
BIC & 0 & 2 & 0 & 5\\
Cross-Validation & 0 & 3 & 0 & 2\\
Elbow Method & 0 & 1 & 1 & 0\\ 
Fixed & 3 & 4 & 0 & 4\\
Gap Statistic & 1 & 0 & 0 & 2\\
Posterior Distribution & 0 & 0 & 0 & 20\\
Silhouette & 3 & 0 & 0 & 0\\
\end{tabular}
\end{table}

\elandscape 

\bibliography{references.bib}